\documentclass{egpubl}
\usepackage{eg2026}
 
\ConferencePaper        % uncomment for (final) Conference Paper
\CGFccby
\usepackage[T1]{fontenc}
\usepackage{dfadobe}  

\usepackage{cite}  % comment out for biblatex with backend=biber
\BibtexOrBiblatex
\electronicVersion
\PrintedOrElectronic
\ifpdf \usepackage[pdftex]{graphicx} \pdfcompresslevel=9
\else \usepackage[dvips]{graphicx} \fi

\usepackage{egweblnk} 
\usepackage[utf8]{inputenc}
\usepackage{amsmath}
\usepackage{dsfont}
\usepackage{caption}
\usepackage{subcaption}
\usepackage{multirow}
\usepackage{siunitx}
\usepackage{xcolor}
\usepackage{enumitem}
\usepackage{amsfonts}
\usepackage{amsmath}
\usepackage{booktabs}
\usepackage{tikz}  % Overlapping figures
\usetikzlibrary{matrix,positioning,arrows.meta}
\tikzset{
  imgarrow/.style={
    line width=0.1pt,
    -{Stealth[length=3pt,width=3pt]}
  }
}
\usepackage[linesnumbered,ruled,vlined]{algorithm2e}
\definecolor{commentcolor}{rgb}{0.3, 0.6, 0.15}
\SetKwComment{TrComment}{$\triangleright$ }{}
\SetCommentSty{mycommfont}
\newcommand\ch[1]{#1}

\renewcommand{\vec}[1]{\boldsymbol{\mathbf{#1}}}
\newcommand{\sethp}{\mathcal{H}_p}

\newcommand{\citep}[1]{\cite{#1}}
\newcommand{\citet}[1]{\cite{#1}}

\newif\ifshowtomention
\showtomentiontrue % Change to \showtomentionfalse to hide TOMENTION  

\usepackage{environ} % To define environments easily

\NewEnviron{TOMENTION}{%
    \ifshowtomention % Check if we should show it
        \begin{itemize}[label=\textbullet]
            \color{blue}% Set color for items
            \BODY % Use BODY to insert content here.
        \end{itemize}
    \fi
}

\usepackage{colortbl}
\usepackage{graphicx}
\usepackage{array}

\newcolumntype{C}{>{\centering\arraybackslash}m{12mm}}

\definecolor{lightgreen}{rgb}{0.85, 0.93, 0.85} % Light green
\definecolor{lightred}{rgb}{1.0, 0.85, 0.85}    % Light red

\graphicspath{{images/}}

\title{Progressively Projected Newton's Method}

\author[J.\,A. Fern{\'a}ndez-Fern{\'a}ndez, F. L{\"o}schner \& J. Bender]
{\parbox{\textwidth}{\centering J.\,A. Fern{\'a}ndez-Fern{\'a}ndez$^{1}$\orcid{0000-0003-4651-7542} 
        , F. L{\"o}schner$^{1}$\orcid{0000-0001-6818-2953}
        and J. Bender$^{1}$\orcid{0000-0002-1908-4027}
        }
        \\
{\parbox{\textwidth}{\centering $^1$RWTH Aachen University, Aachen, Germany
       }
}
}
\begin{document}

% uncomment for using teaser
\teaser{
  \includegraphics[width=\textwidth]{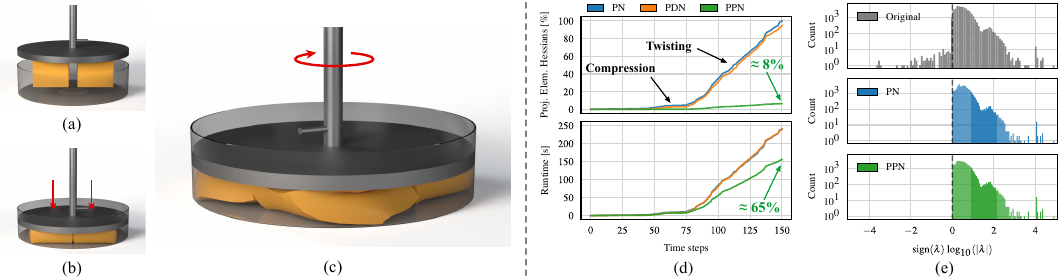}
 \centering
  \caption{%
    \textbf{Press.}
    (a) Four elastic orange boxes drop into a rigid cylinder; 
    (b) a press descends; 
    (c) the press then rotates, forcing the boxes to roll under pressure due to friction.
    (d) \emph{Top}: cumulative count of projected element Hessians, highlighting that PPN projects only 8\,\% of all elements; 
    \emph{bottom}: cumulative runtime highlighting that PPN completes the simulation in 65\,\% of the time required by PN and PDN.  
    (e) Global eigenvalue histogram at time step~125, iteration~4: both PN and PPN handle negative eigenvalues as expected, though not equally; 
    while PN projects all the element Hessians at this iteration, PPN only projects 17.4\,\% of them.}
  \label{fig:teaser}
  % Add back missing spacing between abstract and teaser (no clue why it is gone otherwise)
  \vspace{1.75ex}
}

\maketitle
%-------------------------------------------------------------------------
\begin{abstract}
% \begin{abstract}
    Newton's Method is widely used to find the solution of complex non-linear simulation problems.
    To guarantee a descent direction, it is common practice to clamp the negative eigenvalues of each element Hessian prior to assembly---a strategy known as \emph{Projected Newton} (PN)---but this perturbation often hinders convergence.
    %Recently, \emph{Project-on-Demand Newton} (PDN) was proposed to alleviate this shortcoming by postponing projection until the assembled matrix is detected as indefinite.

    In this work, we observe that projecting only a small subset of element Hessians is sufficient to secure a descent direction.
    Building on this insight, we introduce Progressively Projected Newton (PPN), a novel variant of Newton's Method that uses the current iterate's residual to cheaply determine the subset of element Hessians to project.
    The benefit is twofold: most eigendecompositions are avoided and the global Hessian remains closer to its original form, reducing the number of Newton iterations.

    We compare PPN with PN and Project-on-Demand Newton (PDN) in a comprehensive set of experiments covering contact-free and contact-rich deformables, co-dimensional and rigid-body simulations, and a range of time step sizes, tolerances and resolutions.
    PPN reduces the amount of element projections in dynamic simulations by one order of magnitude while simultaneously improving convergence, consistently being the fastest solver in our benchmark.
% \end{abstract}

\end{abstract}  

%-------------------------------------------------------------------------
%%
%% Sections
\section{Introduction}
\label{sec:introduction}

Robust and efficient simulation of dynamic deformable and rigid objects is a cornerstone of computer graphics, visual effects, and interactive design.  
Many modern simulators perform time-stepping by formulating implicit integration schemes as successive nonlinear optimization problems, each solved with a variant of Newton's method.
When the assembled Hessian of the potential energy is \emph{Symmetric Positive Definite} (SPD), Newton steps are guaranteed to point in descent directions and efficient linear solvers specialized for SPD matrices can be applied, both highly sought-after properties.

In practice, nonlinear materials, large time steps, and other factors frequently render the assembled Hessian indefinite, causing the line search to fail.
As a consequence, practically every nonlinear second-order solver must implement a mechanism to remove indefiniteness from the system.
This stabilization process therefore becomes a fundamental component of this type of solvers, with a strong influence on their robustness and performance.
A widely adopted remedy is \emph{Projected Newton} (PN): an eigendecomposition is performed for every element Hessian and its negative eigenvalues are clamped (or mirrored) before assembly.
Although PN produces an SPD system, it overly distorts the global Hessian and thus slows down convergence by unnecessarily discarding element-local negative curvature information. %, even when the assembled matrix would already be positive definite.
Moreover, projecting all elements imposes the cost of one eigendecomposition per element per Newton step, which can be expensive.
% \eg{
    The root cause of these important drawbacks is the per-element \emph{isolated} view on indefiniteness, which is overly limiting when analyzing the actual indefiniteness of the global system.
% }
While analytic eigenanalysis can alleviate performance concerns, it can be challenging to derive the required expressions when modeling complex effects or materials not yet established in the literature. 
This approach is also incompatible with many simulation frameworks that use automatic or symbolic differentiation for derivative-free modeling and which have to rely on numerical eigendecompositions for flexibility.
%In any case, the convergence rate of Newton remains a concern when using PN.
% A recent analysis quantified these drawbacks~\cite{pitfalls}.

%We propose \emph{Progressively Projected Newton} (PPN), a direct replacement for PN that projects only a subset of element Hessians, and only when proven necessary to obtain descent directions.
We propose \emph{Progressively Projected Newton} (PPN), a direct replacement for PN \ch{for dynamic simulations} that avoids most of the element Hessian projections while ensuring descent directions during the Newton search.
% The key improvement of PPN over PN is that sources of indefiniteness are identified \emph{after} the additive effect of assembly, rather than from an isolated, per-element perspective.
The key improvement of PPN over existing solutions is to identify sources of indefiniteness \emph{after} the additive effect of assembly, rather than from an isolated, per-element perspective.
% PPN produces a global SPD matrix closer to the original than PN, which typically results in fewer Newton iterations, while also avoiding most eigendecompositions.
Each Newton iteration begins with the unaltered Hessian, exactly as in pure Newton's Method.  
If the linear solver exits early due to indefiniteness, PPN projects only those elements whose local residual exceed an adaptive tolerance, updates the global matrix incrementally, and retries the solve.
The tolerance is tightened until a descent direction is reached, then relaxed for the next iteration.
Thus, PPN trades a few inexpensive solver attempts for the elimination of most eigendecompositions and a global Hessian closer to the unmodified one, which often leads to fewer Newton iterations.

PPN behaves exactly like Newton's Method if no projections are required, and degenerates to full PN only in very rare cases.
By projecting only when and where necessary, PPN provides faster convergence than existing methods \ch{in dynamic simulations}, while maintaining the established and reliable robustness of PN.
In typical simulations, PPN omits more than 90\% of element projections, cutting its cost by one order of magnitude, and reducing Newton iterations by up to 50\% compared to PN.
In our implementation, PPN achieves speedups of up to $\times 2.5$ over PN and up to $\times 1.5$ over the best alternative.
% In our implementation, PPN yields speedups between $\times 1.1$ and $\times 1.5$ with respect to state-of-the-art alternatives.

\ch{
    Crucially, PPN is formulated for dynamics, explicitly exploiting the regularization inherent in time integration, most notably the mass matrix.
    This allows it to avoid full-spectrum strategies by recognizing that sources of indefiniteness are highly localized in time and space, effectively rendering spectral verification unnecessary for the vast majority of elements.
    To the best of our knowledge, PPN is the first method to successfully circumvent this global processing requirement.
}

In summary, our contributions are:
\begin{itemize}
    \item Progressively Projected Newton, a novel Newton-type solver that selectively projects element Hessians, drastically cutting eigendecomposition costs and avoiding unnecessary distortions to the global Hessian.
    \item A residual-driven heuristic that ranks elements by their likelihood of contributing negative curvature which reuses information already calculated in the original Newton's Method.
    \item A comprehensive evaluation on contact-free and contact-rich deformable bodies, shells, and rigid body systems.
\end{itemize}

% Overall results
% | Scene               |   % proj avoided |   % Newton iterations reduction |   speedup |
% |---------------------|------------------|---------------------------------|-----------|
% | Press               |               92 |                              14 |     1.540 |
% | U-Turn              |               94 |                              50 |     1.420 |
% | Armadillo slingshot |               97 |                              20 |     1.340 |
% | Twisting cloth      |               90 |                              11 |     1.500 |
% | Armadillo drop      |               94 |                               4 |     1.140 |
% | Tumbler             |               90 |                              12 |     1.120 |

\section{Related Work}
\label{sec:related_work}

We first cover works using optimization-based time integration as our main area of application.
%To ensure convergence of Newton's method in these problems, our approach relies on eigenvalue filtering which we discuss in the following.
We then discuss eigenvalue filtering and briefly introduce ``automatic'' frameworks that use machine-generated derivatives as a prominent use case to apply progressive projections.

\subsection{Optimization-based Time Integration}
Optimization-based time integrators reformulate implicit schemes such as backward Euler as Incremental Potential minimization problems~\citep{ip}, which enables the use of robust optimization methods~\cite{nocedal} in order to advance dynamic simulations in time.
Many works in the computer graphics community adopted this approach~\cite{variational,example_based,fast_mass_spring}.
First-order and quasi-Newton solvers reduce per-iteration cost at the expense of more iterations~\cite{projective_dynamics, admm, primal_dual, Wang16, Liu2017,vbd}.
Despite relatively higher per-iteration costs, second-order approaches are widely used due to strong convergence guarantees~\cite{gast15}.
Applications include frictional contact~\cite{ipc}, cloth and rods~\cite{coipc, thick_shells}, rigid-bodies~\cite{rigid_ipc, affine_bodies}, advanced materials~\cite{micropolar, micropolar_shells} and fluids and granular media~\cite{sph_ipc, duo_ipc}.
For a comprehensive overview of such energy-based models and their coupling we refer readers to the recent multiphysics state-of-the-art report by Holz et al.~\citet{multiphysics_STAR}.
Our proposed Newton-type solver is a drop-in replacement that can improve efficiency in these applications while retaining robustness.

\subsection{Eigenvalue Filtering}
To obtain descent directions during optimization, existing Newton-type solvers typically rely on the global Hessian being SPD.
For most physical models, this is not always the case in practice.
To address this issue, Teran et al.~\citet{teran05} proposed per-element Hessian projection to the cone of SPD matrices by clamping their negative eigenvalues prior to assembly.
This technique, commonly referred to as Projected Newton~\cite{projected_newton}, avoids infeasible global eigendecomposition and facilitates use of linear solvers specific to SPD matrices.
The success of PN motivated a large body of work on efficient per-energy analytic eigenanalysis to avoid expensive numerical eigendecompositions~\cite{smith18, smith19, kim20, eigenanalysis_anisotropic, discrete_bending, arap_chauchy_green, eigenanalysis_collision, eigenanalysis_bending, gipc}.
However, these analytic projections require careful manual modifications of the second-order derivative implementations.
Further approaches for SPD projection include regularization using diagonal matrices~\cite{fu16} or multiples of the mass matrix~\cite{pitfalls}, but these are shown to perform worse than PN in contact-rich scenarios~\cite{pitfalls}.

In the quasistatic setting with strong volume conservation and large initial deformations, eigenvalue \emph{mirroring}~\cite{eigenvalue_mirroring} and variants of \emph{blending}~\cite{eigenvalue_blending,mirroring_trustregion}, as opposed to clamping, have shown to improve convergence.
Unfortunately, as we show in Section~\ref{sec:results}, these results do not directly transfer to dynamic problems.
%In any case, these approaches still require the eigendecomposition of all elements~\todo{verify}, which is our goal to avoid.

A comprehensive study by Longva et al.~\citet{pitfalls} recently demonstrated that unconditional projection slows asymptotic convergence and breaks affine invariance.
Their proposed \emph{Project-on-Demand Newton} (PDN) method performs element projections only when the assembled matrix is detected to be indefinite, which typically occurs far from the solution, recovering Newton-like convergence as the iteration sequence progresses.% while preserving robustness.

The aforementioned strategies share the limitation of acting on \emph{all} elements and \emph{in isolation}, ignoring the additive effect of neighboring contributions, and resulting in ``over-projection''.
Our progressive strategy not only projects on-demand, but also selectively, significantly reducing both the amount of element projections and the distortion imposed on the global Hessian, while still guaranteeing descent directions.

%\todo{Mention that there are other methods than projection, such as regularization but they are less commonly used.}

\subsection{Automatic Frameworks} \label{sec:frameworks}
Recently, a proliferation of frameworks that automate solutions to second-order optimization has occurred in geometry processing and simulation.
These systems rely on machine-generated derivatives and automated evaluation pipelines to tackle complex problems from concise symbolic expressions~\cite{TinyAD, Herholz24, symx}.
They enable rapid and safe composition of solvers and models, thus accelerating research with a measurable impact on the field; several of the referenced works above, for example, were developed on TinyAD~\cite{TinyAD} and SymX~\cite{symx}.

Because semi-analytic projection code remains challenging to generate, these frameworks still rely on numerical eigendecompositions, which is shown to be a very significant cost in our measurements (Section~\ref{sec:results}).
By avoiding most projections, PPN eliminates this bottleneck and further narrows the performance gap with hand-tuned codebases.

%This section introduces our solver and details its algorithmic and implementation components.
%We first review classical Newton's Method and standard eigenvalue filtering.
%We then describe the PPN scheme, introduce the residual-based projection heuristic, and present the full algorithm.
%Finally, we discuss implementation considerations for Hessian updates and linear solves.

\section{Newton's Method}
We seek the configuration $\vec{x}\in\mathbb{R}^{n}$ that minimizes the total potential energy $\Psi(\vec{x})$, which typically aggregates inertial, elastic, frictional and other contributions evaluated on a discretized domain (see, e.g.,~\cite{gast15}).
Applying Newton's method to this optimization problem, at iteration $k$ we solve  
\begin{equation}
    \vec{H}^{k}\,\Delta\vec{x}^{k}=-\vec{g}^{k},
    \label{eq:newton}
\end{equation}
where $\vec{g}^{k}=\nabla_{\vec{x}}\Psi(\vec{x}^{k})$ is the gradient of the energy, $\vec{H}^{k}=\nabla_{\vec{x}}^{2}\Psi(\vec{x}^{k})$ is the Hessian and $\Delta\vec{x}^{k}$ is the Newton step such that $\vec{x}^{k+1} = \vec{x}^k + \Delta\vec{x}^{k}$.
This scheme is applied iteratively until a measure of convergence is fulfilled, e.g. $\vec{g}^{k} \approx \vec{0}$.
Both the global Hessian and gradient are assembled from element contributions $\vec{H}^{k}=\sum_{e}\vec{H}^{e, k}$ and $\vec{g}=\sum_{e}\vec{g}^{e, k}$.
The elements $e$ are typically given by finite elements (e.g. tetrahedra), rigid bodies, particles, collision pairs et cetera.
For clarity we omit the superscript $k$ hereafter.

A non-zero Newton step $\Delta\vec{x}$ is guaranteed to point in a descent direction if $\vec{H}$ is SPD locally at the current iterate, that is,
\begin{equation}
    \vec{r}^T\vec{H}\,\vec{r}>0\,,\quad\forall\,\vec{r}\neq\vec{0}\,,
    \label{eq:spd}
\end{equation}  
which implies that $\Psi$ features strictly positive curvature locally in every direction, i.e.\ it is locally \emph{convex}.
The connection between an SPD Hessian and a descent direction can be shown by multiplying both sides of Eq.~\eqref{eq:newton} by $\Delta\vec{x}^T$
\begin{equation}
    \Delta\vec{x}^T\vec{g}=-\Delta\vec{x}^T\vec{H}\,\Delta\vec{x}\,.
\end{equation}  
If $\vec{H}$ is SPD, the right-hand side is strictly negative, hence $\Delta\vec{x}^T\vec{g}<0$ \citep{nocedal}.

In practice however, $\vec{H}$ is often indefinite (see, e.g.,~\cite{dynamic_deformables}).
PN remedies this by filtering (e.g. clamping) the negative eigenvalues of the element Hessians prior to assembly.
Consider the eigendecomposition of the Hessian of element $e$
\begin{equation}
    \vec{H}^{e}=\vec{Q}^{e}\,\vec{D}^{e}\bigl(\vec{Q}^{e}\bigr)^T,
\end{equation}
where the columns of $\vec{Q}^{e}$ are the eigenvectors of $\vec{H}^{e}$ and $\vec{D}^{e}$ is a diagonal matrix of the corresponding eigenvalues.
Applying clamping, the respective SPD projected element Hessian is then $\widehat{\vec{H}}^{e}=\vec{Q}^{e}\,\widehat{\vec{D}}^{e}\bigl(\vec{Q}^{e}\bigr)^T$, for $\widehat{D}^{e}_{ii}=\max(D^{e}_{ii},\varepsilon)$ with $\varepsilon>0$.
As a sum of SPD matrices is SPD, which holds for the assembled global matrix.
%As a result, the global matrix has two very important properties: it is relatively close to the original Hessian, which provides good convergence, and it guarantees descent directions.

However, projecting all the element Hessians is unnecessary.
Consider element matrices $A$ and $B$, and the global matrix $P$:
\[
    A=\begin{pmatrix}-1 & 0\\ 0 & 2\end{pmatrix},\quad
    B=\begin{pmatrix} 2 & 0\\ 0 & 2\end{pmatrix},\quad
    P=A+B=\begin{pmatrix} 1 & 0\\ 0 & 4\end{pmatrix}.
\]  
$A$ is indefinite with eigenvalues $\{-1,2\}$, $B$ is SPD with eigenvalues $\{2,2\}$, yet the assembled $P$ is SPD with eigenvalues $\{1,4\}$.  
Projecting $A$ would alter the global matrix unnecessarily and likely deteriorate convergence of Newton's method as shown by Longva et al.~\citet{pitfalls}.
Project-on-Demand Newton addresses this issue by projecting all element Hessians only if the global matrix is proven indefinite, generally improving convergence over PN.

\section{Progressively Projected Newton's Method}
\label{sec:method}

In this section we introduce our method, PPN, starting with a motivating example.
Consider a third element matrix added to the previous ones:
\[
    C=\begin{pmatrix} -10 & 0\\ 0 & 1\end{pmatrix},\quad
    Q=A+B+C=\begin{pmatrix} -9 & 0\\ 0 & 5\end{pmatrix}.
\]
In this case, $Q$ is indefinite with eigenvalues $\{-9,5\}$.
However, only projecting $C$ suffices to obtain an SPD approximation, demonstrating that full projection, even when conditionally triggered as PDN does, is still unnecessary.
This \emph{element-selective projection} is the core idea of PPN: the global Hessian $\vec{H}$ is built from two disjoint sets of element Hessians, the projected set $\sethp$ and the unprojected set $\mathcal{H}_u$.
The goal then is to keep $\lvert\mathcal{H}_{p}\rvert$ minimal while ensuring that the assembled Hessian yields a descent direction at all times during Newton iterations.
% The benefit is twofold: unnecessary (and potentially expensive) element projections are avoided, and the global Hessian is kept closer to the true Hessian, generally improving convergence.

In practice, however, identifying which element Hessians to project is not as straightforward as in the example above.
Assembly scatters element contributions, making global eigenvalue recovery intractable, and building $\sethp$ based on exact global eigenvalues would be prohibitively expensive.
Instead, we propose a cheap post-assembly proxy for the \emph{likelihood} that an element is the source of global indefiniteness.
We pair this prediction with a robust global indefiniteness check to adapt $\sethp$ as needed.
Both the predictor and the indefiniteness check derive efficiently from information already available during the Newton minimization process.

% Figure: Heuristic Validation
\begin{figure}
    \centering
    \includegraphics[width=\columnwidth,trim={0 0 0 0},clip]{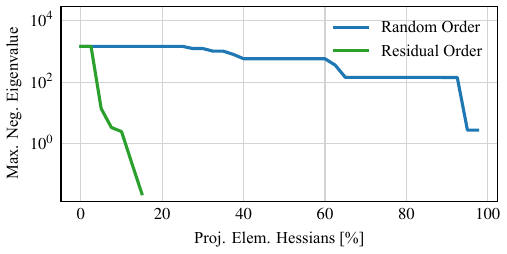}
    \caption{\ch{
        Largest global negative eigenvalue as element Hessians are progressively projected for the iteration shown in Figure~\ref{fig:teaser} (d) and (e).
        In blue, randomized element order. In green, elements ordered by the local assembled residual from largest to smallest.
    }}
    \label{fig:heuristic_validation}
\end{figure}

\paragraph*{Projection heuristic} %
We motivate a residual-based heuristic to rank elements by their likelihood of contributing to global indefiniteness. 
Let $\vec{x}^{\star}$ be a strict local minimizer with $\vec{g}(\vec{x}^{\star})=\vec{0}$ and $\Psi(\vec{x}^{\star}+\vec{\epsilon})>\Psi(\vec{x}^{\star})$ for small $\vec{\epsilon}$.
By Eq.~\eqref{eq:spd}, $\vec{H}(\vec{x}^{\star})$ is SPD, and by continuity it remains (semi)definite in a neighborhood of $\vec{x}^{\star}$. 
Hence, as $\vec{x}\!\to\!\vec{x}^{\star}$, $\vec{g}\!\to\!\vec{0}$ and indefiniteness in $\vec{H}$ diminishes.
However, convergence does not occur uniformly across the domain.
We observed that elements in regions with lower assembled residual are less likely to induce indefiniteness in the global system, even if left unprojected.
\ch{
    Figure~\ref{fig:heuristic_validation} empirically supports this hypothesis: while a random ordering requires the projection of almost all elements to recover global positive definiteness, projecting in order of larger assembled residual first isolates the indefinite contributions rapidly.
    Elements in low-residual regions are more likely to either already have all-positive eigenvalues or their minor indefinite spectra cannot overturn stronger SPD contributions, such as the influence of the mass matrix.
    As demonstrated in Section~\ref{sec:results}, this correlation is robust and consistent across a diverse set of scenarios, material models, and discretization parameters.
}

Motivated by this, we prioritize projecting element Hessians in regions with larger assembled residual and expand as necessary.
Specifically, we introduce a projection tolerance $\delta$ and add $\vec{H}^e$ to $\mathcal{H}_{p}$ when the element-local residual of $e$ is larger than $\delta$, that is, $\lVert\vec{S}^{e} \vec{g}\rVert _{\infty} > \delta$, where $\vec{S}^{e}$ is a selection matrix that extracts the assembled entries affected by element $e$.
\ch{
    The $\infty$-norm captures the largest component of the local residual, acting as a selector of all elements incident to nodes with large residuals, aligning with the concept of domain partitioning.
    The gradient, or residual, relates directly to stationarity where the system is known to be SPD while other progress metrics, such as the Newton step or iteration decrements, depend on the optimization procedure.
}
The tolerance $\delta$ is adapted over the course of the Newton iterations by a tightening factor $\alpha\in(0,1)$ and a release factor $\beta\geq1$.
The former is applied when indefiniteness is detected in the global Hessian, and the latter after a successful step is taken.
As a result, we obtain an effective partition using already calculated values at no extra cost.
\ch{This parametrization exhibits low sensitivity, yielding consistent performance gains across broad ranges of values.}
Values of $\alpha=0.5$ and $\beta=2$ are chosen as safe defaults by ablation tests in Section~\ref{sec:results}.
\ch{To guarantee termination and bound the worst-case runtime, we fall back to PN and project all elements if $\delta$ drops below a small threshold $\varepsilon_{\delta} = 10^{-12}$.}

\begin{algorithm}
\caption{Progressively Projected Newton (PPN)}
\label{alg:ppn}
\DontPrintSemicolon
$\delta \gets \infty$ \\
\While{not converged}{
  Assemble unprojected $\mathbf{H}$ and $\mathbf{g}$ \\
  $(\Delta \mathbf{x},\,\textit{ind}) \gets \textsc{SolveSPD}(\mathbf{H},-\mathbf{g})$ \\
  \While{\textit{ind}}{
    \If{$\delta=\infty$}{$\delta \gets \alpha\,\lVert \mathbf{g}\rVert_{\infty}$}
    \ForEach{$e$}{%
        \If{$(\lVert \mathbf{S}^{e}\,\mathbf{g}\rVert_{\infty} > \delta) \; \textup{\textbf{or}} \; (\delta\le\varepsilon_{\delta})$} {
            \textsc{ProjectToPD}$(\mathbf{H},e)$ \TrComment*[r]{In-place update}
        }
    }
    $(\Delta \mathbf{x},\,\textit{ind}) \gets \textsc{SolveSPD}(\mathbf{H},-\mathbf{g})$ \\
    \If{\textit{ind}}{
        $\delta \gets \alpha\,\delta$ \TrComment*[r]{Tighten projection}
        }  
  }
  $\delta \gets \beta\,\delta$ \TrComment*[r]{Relax projection}
  $\gamma \gets \textsc{LineSearch}(\Delta \mathbf{x})$ \\
  $\mathbf{x} \gets \mathbf{x} + \gamma\,\Delta \mathbf{x}$ \\
}
\end{algorithm}

\paragraph*{Algorithm} %
Our method is outlined in Algorithm~\ref{alg:ppn}.
%Before the solve, $\delta$ is set to infinity (line 1) to project no elements.
The standard Newton's Method logic with an SPD linear system solver is unmodified except for the inner PPN projection logic (lines 5 to 13).
Note that the linear solver must report whether indefiniteness was encountered in addition to the solution $\Delta \vec{x}$.
%Element-wise projections to PD are progressively done until the linear system solver does not encounter indefiniteness (lines 5 to 11).
The first time projection is needed, $\delta$ is initialized in relation to the largest absolute value of the current residual (lines 6 and 7).
%The set $\sethp$ is updated using the new $\delta$ and $\vec{H}$ is updated with the newly projected element Hessians (line 8).
%The updated linear system is solved (line 9).
If the system still cannot be solved, $\delta$ is reduced (lines 12 and 13).
Once the liner solve is successful, $\delta$ is increased for the next Newton iteration (line 14).
\ch{Eigenvalue clamping~\cite{teran05} is used for the projection of element Hessians.}

\subsection{Implementation}
PPN integrates naturally into existing PN pipelines, requiring only two operations to be done efficiently: incremental global Hessian updates, and an SPD linear solver that exits early upon indefiniteness.

\paragraph*{Incremental Hessian updates} %
When element $e$ moves from $\mathcal{H}_{u}$ to $\mathcal{H}_{p}$ the update can be done by assembling the difference
\begin{equation}
    \Delta\vec{H}^{e}=\widehat{\vec{H}}^{e}-\vec{H}^{e}
\end{equation}
into the global matrix in place, which can reuse existing assembly routines.
Importantly, this operation does not change the sparsity of $\vec{H}$, which should make updates faster than the original assembly.

\paragraph*{Linear Solvers} %
In this work we consider Preconditioned Conjugate Gradient (PCG) and Cholesky factorization solves (LLT), which can both exit early on indefiniteness for significantly lower cost than the total cost of the linear solve.

The numerical factorization of LLT exits on the first negative pivot encountered.
Since the expensive symbolic analysis can be reused as long as the sparsity pattern does not change, failing is amortized with the eventual successful solve.
% When paired with explicit checks for SPDness, such as a Cholesky factorization, PPN produces SPD global Hessians.

In the case of PCG, we monitor its intermediate value $\vec{d}^T\vec{H}\vec{d}$ for each CG search direction $\vec{d}$.
If a direction of negative curvature is encountered, indefiniteness is confirmed and the linear solver is stopped.
Thus, when equipped with (P)CG, PPN draws parallels with the ``Newton-CG'' method: instead of using the last valid intermediate solution of CG as the Newton step (which might be very inaccurate), we restart the CG solve with more projections applied.
Even if we do not eliminate all indefiniteness from the global Hessian, as long as CG does not encounter a negative search direction, the resulting intermediate solution is guaranteed to be a descent direction as in Newton-CG~\cite{nocedal}.
In our experiments, warm-starting subsequent PCG calls with the last descent direction yielded worse results than simply starting every solve with the zero vector, hence we use the latter approach.
%This check ensures that all the steps are taken in definite directions, which suffices to obtain a descent direction.%, not that $\vec{H}$ is strictly SPD.
%In any case, in our extensive testing, this procedure never failed to produce descent directions in the Newton search.

\paragraph*{Multithreading}%
While our method adds a few additional steps, it does not introduce thread load imbalance and remains friendly to multithreading:
the unprojected Hessian is built (all threads, unrelated to PPN),
the linear solver is run (all threads, also unrelated to PPN),
and, if indefiniteness is detected, a selection of elements is projected in place (all threads).
Section~\ref{sec:results} demonstrates our method's performance on a high-core-count workstation CPU.

% Notes:
%  - Do we include a PN simulation where projection is done at the time of building? This can be an easy complain.
%  - We do not project the next iteration after a backstep is taken in the line search as we found it too punishing and resulted in PDN leaning even more into PN.
%  - We use the BCRS facilities included in STARK/SymX. Element updates are carried in parallel using a mutex per block-row.
%  - We use intersection detection instead of CCD.

\section{Results}
\label{sec:results}
In this section, we present a comprehensive suite of experiments to compare PPN with PN and PDN across a variety of simulations.
Before that, we describe the hardware, software, and models used in our experiments, followed by an ablation study on PPN's parameters.

\subsection{Experimental Setup}
\paragraph*{Hardware and software.}
%All experiments are conducted on a workstation equipped with an \SI{3.60}{\giga\hertz} AMD Ryzen\textsuperscript{\texttrademark} Threadripper\textsuperscript{\texttrademark} PRO 5975WX processor (32 cores, 64 threads) and \SI{256}{\gibi\byte} of RAM.
All experiments are conducted on a workstation equipped with a \SI{3.60}{\giga\hertz} AMD Ryzen Threadripper PRO 5975WX processor (32 cores, 64 threads) and \SI{256}{\giga\byte} of RAM.
Code is compiled with {gcc 12.2} and built on top of the open-source \textsc{stark} simulation framework~\cite{stark}. 
We use the framework's built-in 3$\times$3 Blocked Diagonal PCG solver and {Intel MKL 2025} for Cholesky factorization.
{Eigen 3.4} handles all other linear algebra operations, including eigendecompositions, which we measure to be on average $\times$1.53 faster than MKL's ones for matrices of size 15$\times$15 and smaller.

\begin{figure}
    \centering
    \includegraphics[width=\columnwidth,trim={0 430 0 350},clip]{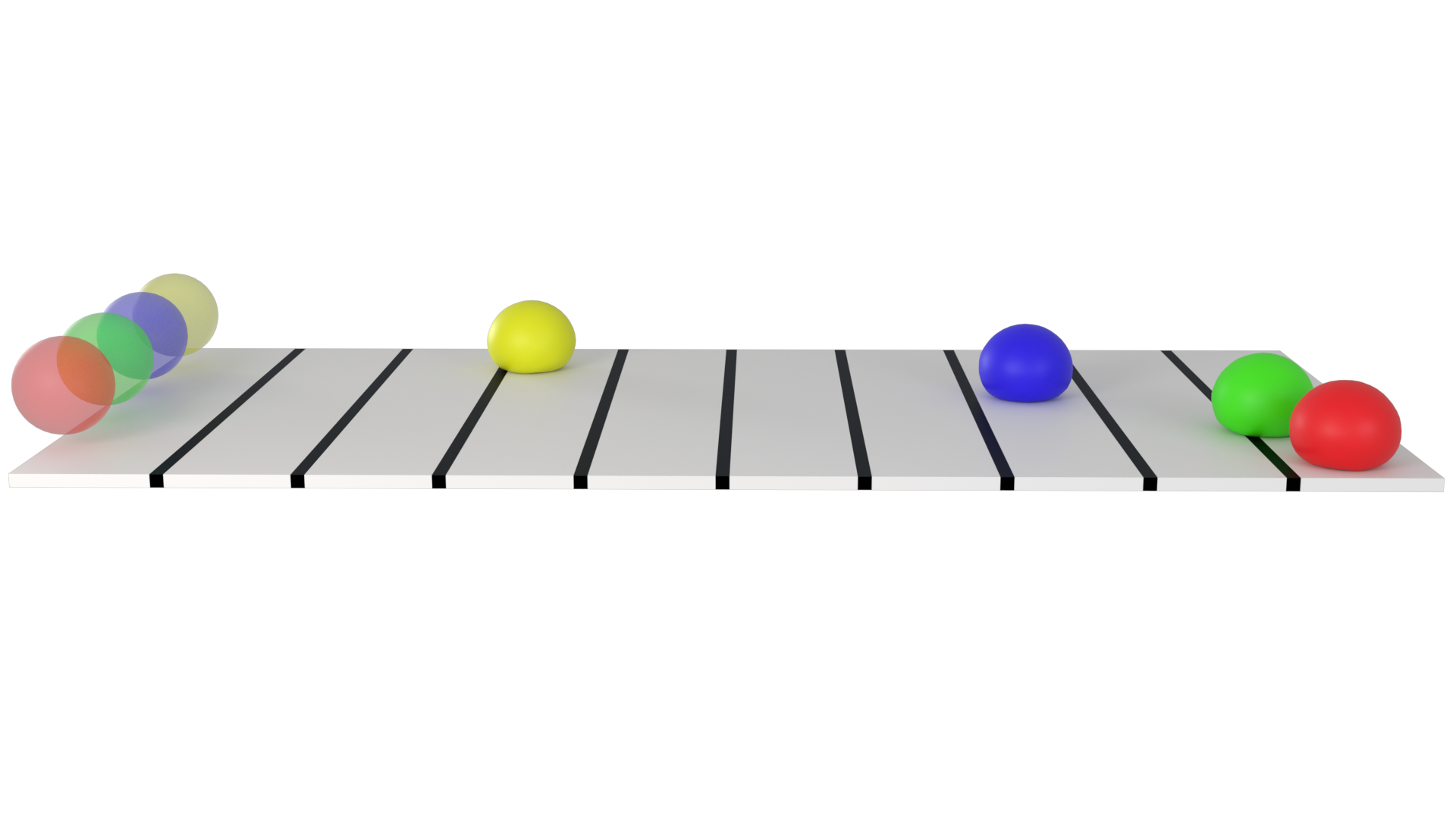}
    \caption{
        \textbf{Rolling sphere.}
        A deformable sphere with initial horizontal velocity rolls on a flat surface.
        Different Newton step velocity tolerance are used:
        $10^{-1}\SI{}{\meter\per\second}$ (yellow), $10^{-2}\SI{}{\meter\per\second}$ (blue), $10^{-3}\SI{}{\meter\per\second}$ (green), $10^{-4}\SI{}{\meter\per\second}$ (red).
        }
    \label{fig:sphere_roll}
\end{figure}

\paragraph*{Time stepping and tolerances.}
We use the backward Euler scheme for time stepping and, unless otherwise stated, a time step size of $\Delta t = 1/30\,$\SI{}{\second}.
As stopping tolerance for Newton's method we check if the velocity step infinity-norm $\Delta t^{-1}\lVert \Delta \vec{x}\rVert _{\infty}$ falls below $10^{-3}\,\si{\meter\per\second}$.
The choice of tolerance greatly influences the number of Newton iterations.
The experiment shown in Fig.~\ref{fig:sphere_roll} justifies our choice: a tolerance of $10^{-2}\,\si{\meter\per\second}$ or coarser causes outcome-altering energy losses, while tolerances of $10^{-3}\,\si{\meter\per\second}$ and $10^{-4}\,\si{\meter\per\second}$ lie much closer.
For consistent comparisons and to avoid bias from inexactness, we verify convergence across solvers with a final fully projected Hessian solve (as PN would) via LLT factorization.
Because such a validation solve is not typically performed in production, we exclude its cost from all timing measurements.  
PCG uses a relative residual tolerance of $\lVert \vec{r} \rVert \lVert \vec{r}^0 \rVert^{-1} = 10^{-4}$.
Element eigenvalues are clamped to $\varepsilon=10^{-8}$.  
For PDN, we adopt the countdown of~4 suggested in the original paper, which also yielded the best results in our “Press'' benchmark.

\paragraph*{Boundary conditions and materials.}
Dirichlet constraints are enforced using penalty potentials.  
All elastic solids employ the Neo-Hookean material in 2D and the Stable Neo-Hookean~\cite{smith18} model in 3D.
Frictional contact uses the IPC~\cite{ipc} potentials with backtracking line search for sufficient descent and intersection-based filtering. 
We use the rigid body inertial potential by Macklin et al.~\citet{primal_dual}.
A list of material parameters, mesh sizes, and time step settings is provided in Table~\ref{tab:sim_params}.

%\paragraph*{Runtime considerations.}
%For a clearer presentation of results, we separate Hessian evaluation (\texttt{Evaluate}), projection (\texttt{Project to PD}), and assembly (\texttt{Assembly}), rather than performing them concurrently on each thread-element.
%Regardless, the sub-optimality introduced (for all methods) is measured at under \todo{XX}\%.

% Figure: Eigen Decomposition Bench

% Figure: Rolling sphere

\subsection{Experiments}
\begin{figure}[t]
    \centering
    \includegraphics[width=\columnwidth,trim={0 22 0 0},clip]{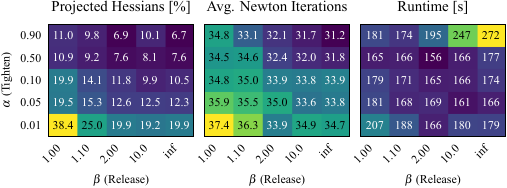}\\
    \includegraphics[width=\columnwidth,trim={0 0 0 9},clip]{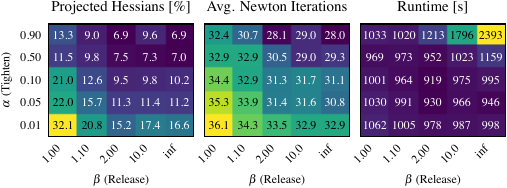}\\
    \caption{
            Ablation test for the parametrization of PPN in the Press scene using the iterative PCG linear solver (top) and the direct LLT solver (bottom).
            Color scale is independent per table.
    }
    \label{fig:ablation}
\end{figure}

\begin{figure}[t]
    \centering
    \includegraphics[width=\columnwidth,trim={0 22 0 0},clip]{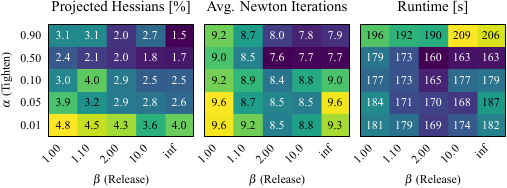}\\
    \includegraphics[width=\columnwidth,trim={0 0 0 9},clip]{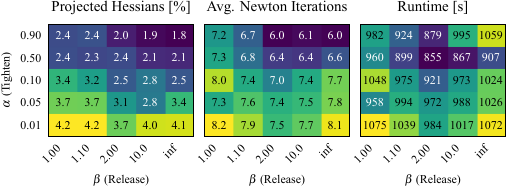}\\
    \caption{
        \ch{
            Ablation test for the parametrization of PPN in the Armadillo Slingshot scene using the iterative PCG linear solver (top) and the direct LLT solver (bottom).
            Color scale is independent per table.
        }
    }
    \label{fig:ablation_armadillo}
\end{figure}

\begin{figure}
    \centering
    % ── big runtime / overview plot ──────────────────────────────────
    \includegraphics[width=\columnwidth]{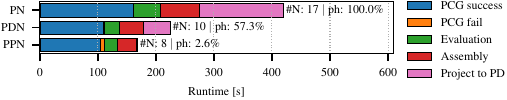}
  
    \medskip  % vertical gap
  
    % ── two snapshots + centred arrow + labels -----------------------
    \begin{tikzpicture}[inner sep=0]
      \sffamily
      % left (initial) image
      \node (A)
        {\reflectbox{%
            \includegraphics[width=0.504\columnwidth,
                             trim={800 350 0 400},clip]{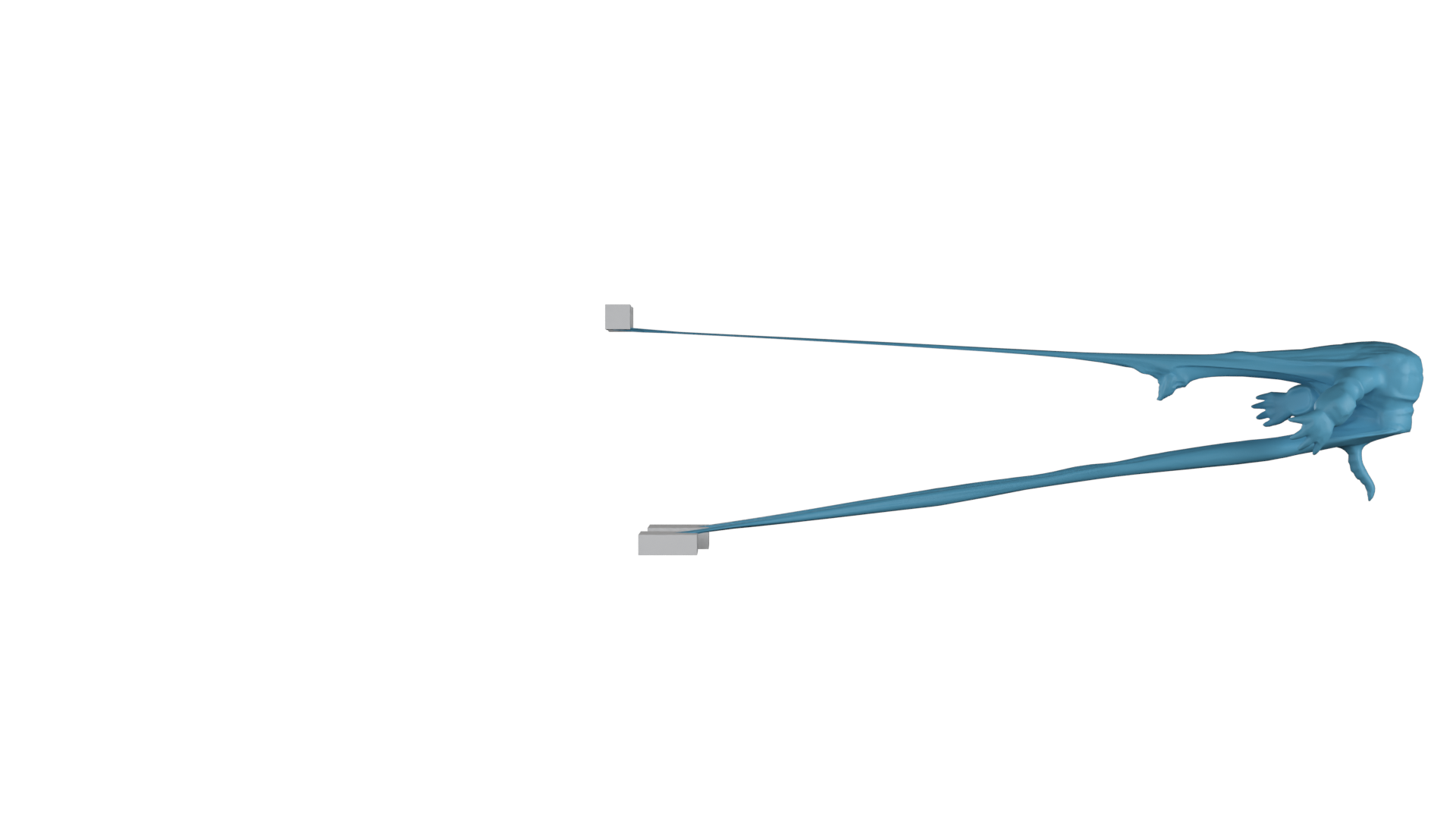}}};
  
      % right (final) image
      \node[right=15pt of A] (B)
        {\reflectbox{%
            \includegraphics[width=0.42\columnwidth,
                             trim={0 350 1000 400},clip]{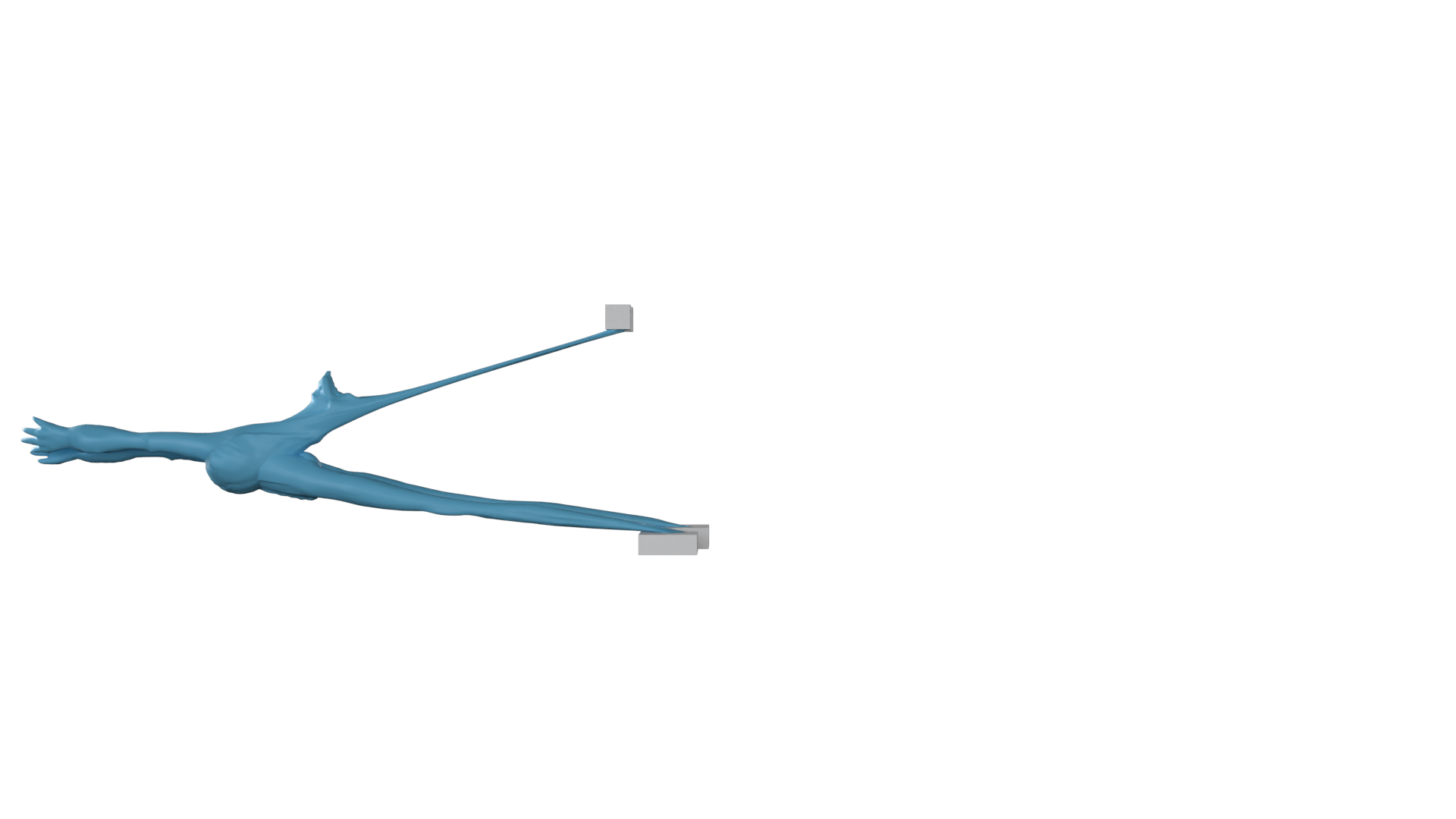}}};
  
      % arrow centred between the two pictures
      \draw[imgarrow] (A.east) -- (B.west);
  
      % column labels
      \node[below=6pt of A.south, anchor=north] {\scriptsize Pull};
      \node[below=6pt of B.south, anchor=north] {\scriptsize Release};
    \end{tikzpicture}
  
    \caption{\textbf{Armadillo slingshot}.
        An elastic armadillo is pulled and then released.
        Comparison between solvers on top, largest deformation states below.
        }
    \label{fig:slingshot}
  \end{figure}

\paragraph*{Ablation}
\ch{
    We systematically analyze the sensitivity of PPN to the tightening ($\alpha$) and release ($\beta$) rates.
    Figures~\ref{fig:ablation} and~\ref{fig:ablation_armadillo} map the performance landscape for the ``Press'' (Fig.~\ref{fig:teaser}) and ``Armadillo Slingshot'' (Fig.~\ref{fig:slingshot}) scenes, covering both iterative (PCG) and direct (LLT) linear solvers.
    These scenes provide complementary stress-tests: ``Press'' is dominated by compression and friction-induced indefiniteness, whereas ``Armadillo Slingshot'' characterizes a regime of extreme tensile strains without contact.
    In the aggressive limit ($\alpha = 0.9, \beta \to \infty$), the method successfully minimizes projection counts ($<7\%$ and $<2\%$ respectively), but total runtime degrades due to the overhead of recovering from failed indefinite linear solves.
    Conversely, conservative settings ($\alpha = 0.01, \beta = 1.0$), result in excessive projection and increased Newton iterations, which is also slow.
    We adopt a power-of-two scaling ($\alpha = 0.5$ and $\beta = 2.0$) as our standard parametrization, owing to its optimal or near-optimal performance.
    The method exhibits low sensitivity around this optimum, allowing us to apply these parameters across all scenes without individual fine-tuning.
}

\begin{figure}
    \centering
    \includegraphics[width=\columnwidth,trim={0 0 0 0},clip]{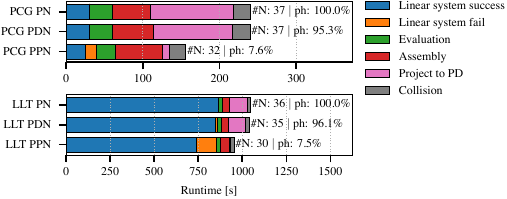}
    \caption{
        Runtime breakdown for the chosen parametrization of PPN in the ``Press'' scene using PCG (top) and LLT (bottom) linear solvers.
        The average number of Newton iterations (\texttt{\#N}) and the percentage of projected element Hessians (\texttt{ph}) are shown at the right of each bar.
        }
    \label{fig:ref_runtimes}
\end{figure}

Next, we compare PPN with PN and PDN, using both PCG and LLT, in Fig.~\ref{fig:ref_runtimes}.
We observe that PDN largely resorts to PN, revealing that only a few steps encountered zero global indefiniteness.
Even in this stress conditions, the adaptive nature of PPN avoids over 92\% of element projections and reduces Newton iterations by 14\%.
Runtime gains with LLT are modest (5.2\%), as the direct solver dominates total cost, but with PCG, we observe a speedup of $\times$1.5 compared to PN and PDN.
See Fig.~\ref{fig:teaser} (d) for a visualization of these reductions over the time steps, and Fig.~\ref{fig:teaser} (e) for the effect on negative eigenvalues of the three methods.
All following experiments use PCG as the linear solver.

%We also solve this problem using eigenvalue mirroring for all three solvers.
Eigenvalue mirroring~\cite{eigenvalue_mirroring} was originally introduced specifically for quasistatic problems where strong indefiniteness is not counteracted by e.g. the mass matrix.
For completeness, we apply mirroring to the ``Press'' experiment for all three solvers, however, results are poor in this elastodynamics with contact simulation: Mirroring consistently performs worse than clamping, needing an average of 51, 52, and 45 Newton iterations for PN, PDN, and PPN respectively, an increase of about 40\% across all solvers.
Based on this result, we apply clamping for all further dynamic experiments if not otherwise specified.

% Ablation Color Map

% Ablation Color Map: Armadillo

% Ref runtime

%|         |   runtime_actual_total |   n_total_newton |   total_ratio_projected_hessians |
%|---------|------------------------|------------------|----------------------------------|
%| PCG PN  |                240.558 |         5721.000 |                            1.000 |
%| PCG PDN |                239.395 |         5666.000 |                            0.953 |
%| PCG PPN |                155.700 |         4894.000 |                            0.076 |
%
%| LLT PN  |               1047.456 |         5444.000 |                            1.000 |
%| LLT PDN |               1038.769 |         5312.000 |                            0.961 |
%| LLT PPN |                952.416 |         4608.000 |                            0.075 |

\begin{figure}
    \centering
    \includegraphics[width=\columnwidth,trim={0 0 0 0},clip]{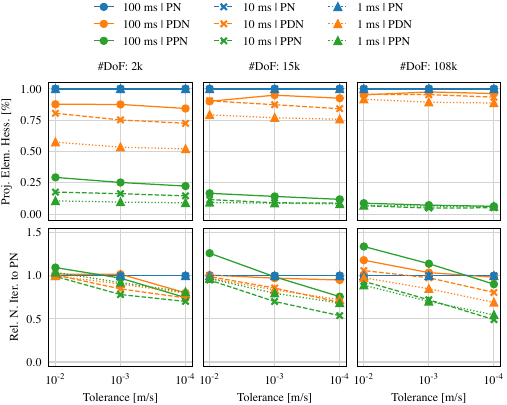}
    \caption{
        Number of projected Hessians (top) and Newton iterations (bottom) in relation to PN for all solvers in different parametrizations of the ``Press'' scene:
        mesh resolution (columns), time step size (marker shape), solver type (color) and tolerance (x-axis).
        }
    \label{fig:operational_range}
\end{figure}

\paragraph*{Resolution, Time Step Size and Tolerance}
Fig.~\ref{fig:operational_range} compares all solvers on the ``Press'' scene across different resolutions (2k, 15k, 108k degrees of freedom), time step sizes (100, 10, 1 ms) and tolerances ($10^{-2}$, $10^{-3}$, $10^{-4}$ $\SI{}{\meter\per\second}$).
PPN solves all instances by projecting only a fraction of the elements (between 30\% and 5\%), correlating positively with finer resolutions, smaller time steps, and tighter tolerances:
Finer resolutions localize sources of indefiniteness more effectively, smaller time steps magnify the regularizing effect of the mass matrix, and stricter tolerances extend the Newton iteration sequence with steps where most of the domain has locally converged.
PPN also achieves consistently fewer Newton iterations and lower runtimes, with the exception of the very large time step of $\Delta t = 100$ ms, where both PDN and PPN struggle.
%For this case, PN delivers the best results, indicating that in the absence of a strong mass matrix, unconditional projection is a better option than on-demand.
Nevertheless, PPN performs strongly, completing the entire benchmark using 27.3\% fewer Newton iterations than PN and offering a speedup of $\times$1.93.
In contrast, PDN takes 7.7\% fewer Newton iterations and is just 7\% faster when compared to PN.

% Figure: Operational range

% [PDN/PN] Proj ratio:	 0.839 | [0.521, 0.976]
% [PDN/PN] #Newton:		 0.916 | [0.691, 1.182]
% [PDN/PN] Runtime:		 0.871 | [0.640, 1.128]
% [PDN/PN] Sum Runtime:	 0.775
% [PDN/PN] Sum Runtime dt=10.0ms & tol=0.001: 0.923
% [PDN/PN] Sum Runtime dt=10.0ms & tol=0.001: 0.930

% [PPN/PN] Proj ratio:	 0.116 | [0.049, 0.294]
% [PPN/PN] #Newton:		 0.867 | [0.492, 1.341]
% [PPN/PN] Runtime:		 0.781 | [0.312, 1.368]
% [PDN/PN] Sum Runtime:	 0.477
% [PPN/PN] Sum Runtime dt=10.0ms & tol=0.001: 0.727
% [PPN/PN] Sum Runtime dt=10.0ms & tol=0.001: 0.491

\begin{figure}
\centering
% ── big plot on top ─────────────────────────────────────────
\includegraphics[width=\columnwidth]{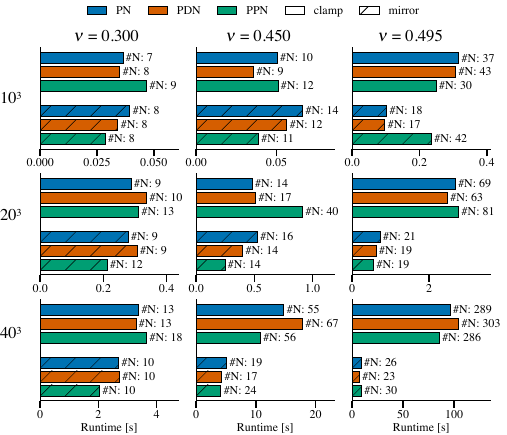} \\
\vspace{1.0em}%
% ── 2 images + centred arrow (TikZ) ─────────────────────────
%    • first node holds the left image
%    • second node holds the right image
%    • the draw command connects their *bounding boxes*, so the
%      arrow is perfectly centred vertically.
\begin{tikzpicture}[baseline=(A.base)]
    \sffamily
    % left image + label
    \node[inner sep=0] (A)
        {\includegraphics[width=0.35\columnwidth,
                        trim={0 250 0 250},clip]{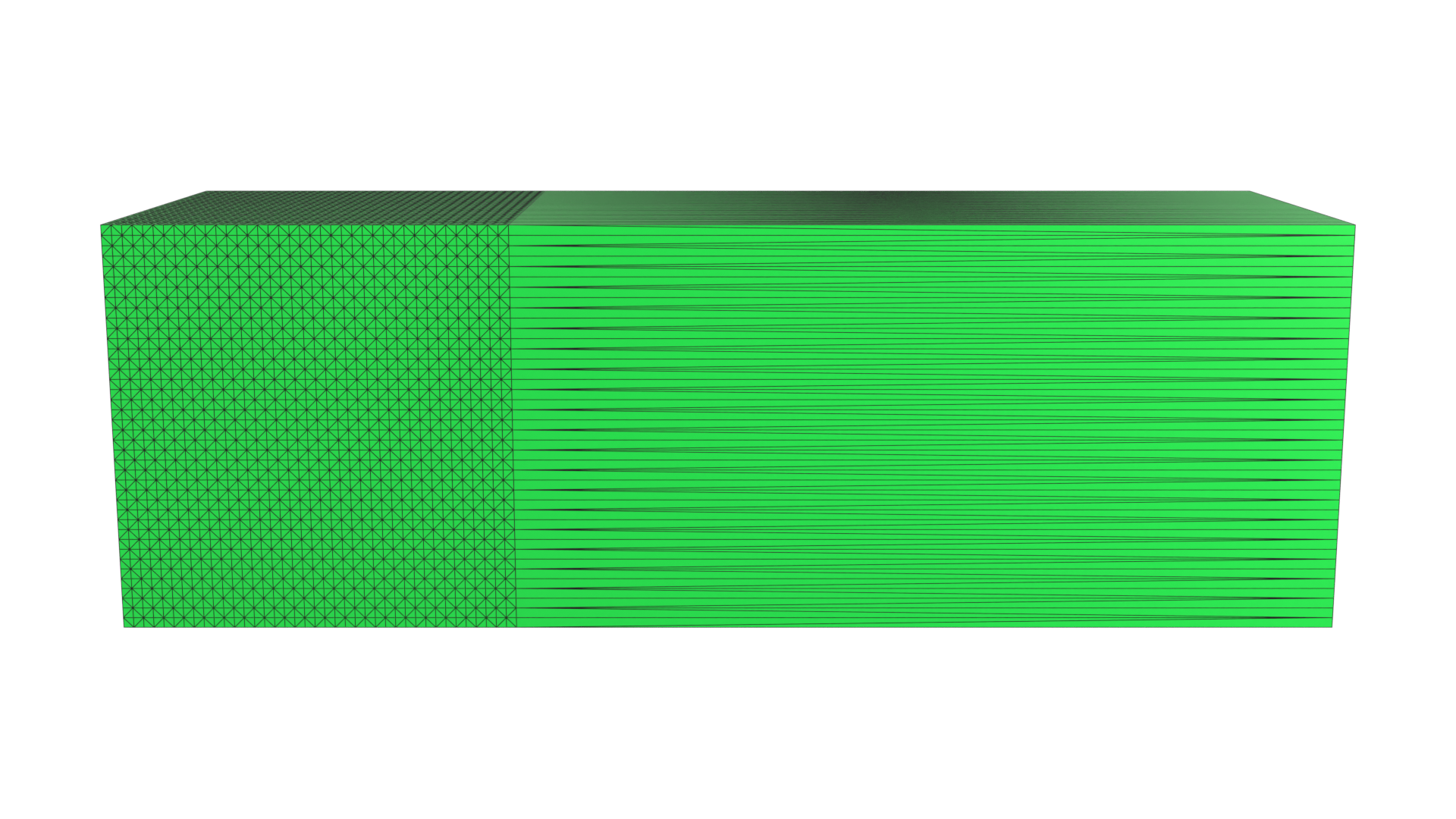}};
    \node[below=2pt of A, font=\scriptsize] {Initial State};
    % right image + label
    \node[inner sep=0, right=8mm of A] (B)
        {\includegraphics[width=0.35\columnwidth,
                        trim={0 250 0 250},clip]{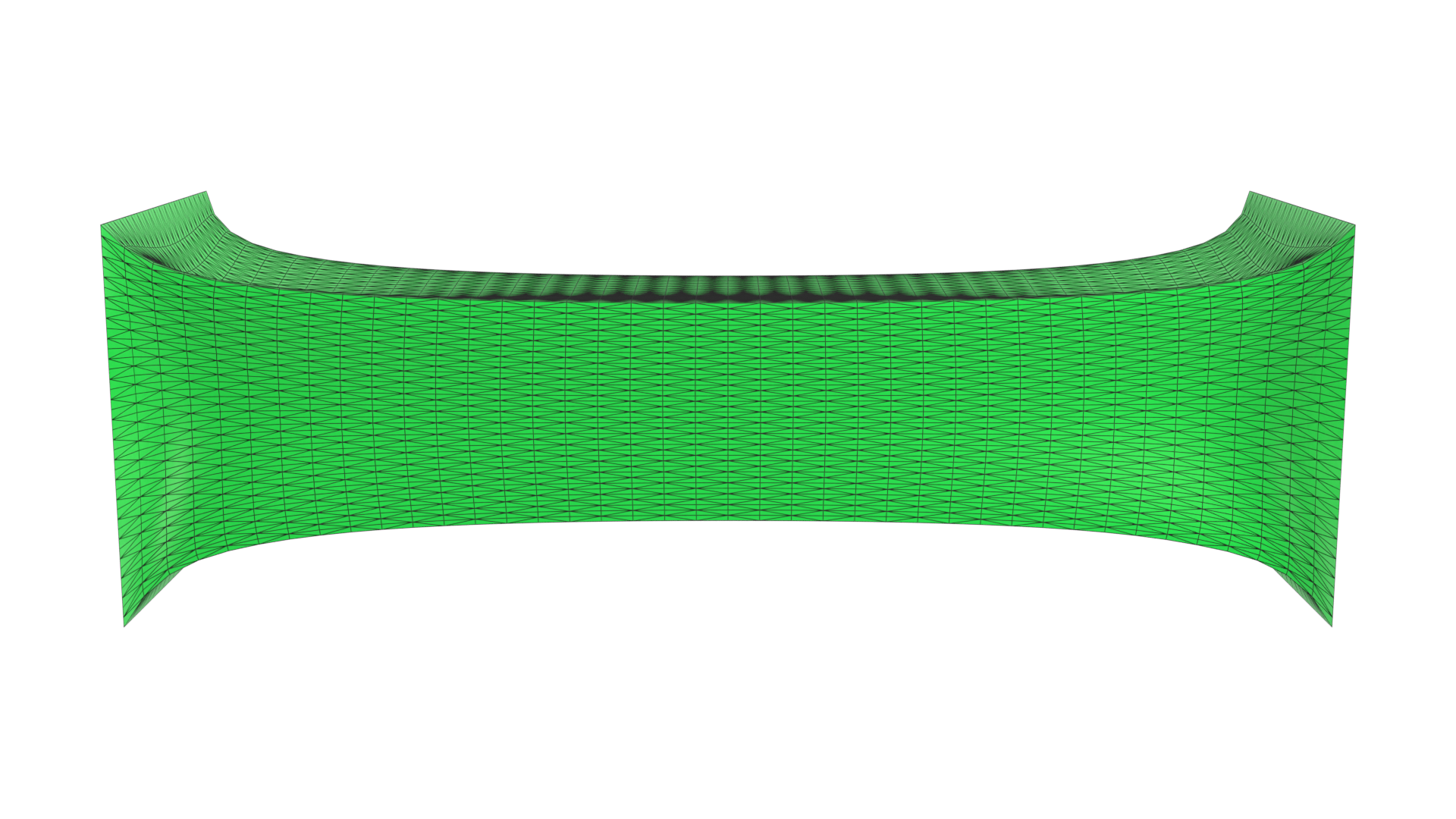}};
    \node[below=2pt of B, font=\scriptsize] {Converged};
    % arrow between the two
    \draw[imgarrow] (A.east) -- (B.west);
\end{tikzpicture}%
\caption{
    \textbf{Quasistatic extrusion}.
    An elastic box is stretched by twice its size in a quasistatic setting using different resolutions (rows) and Poisson's ratios (columns).
    Solvers use different colors, and eigenvalue filtering different shading: solid for clamping and hatched for mirroring.
    The number of Newton iterations is shown at the right of each bar.
    }
\label{fig:quasistatics}
\end{figure}

\paragraph*{Quasistatic Simulation}
We compare the three solvers using eigenvalue clamping and mirroring on a quasistatic problem involving a large initial deformation for various resolutions and Poisson ratios (Fig.~\ref{fig:quasistatics}).
We use a Newton step stopping criteria of 0.1\% of the domain's size.
In line with Chen et al.~\citet{eigenvalue_mirroring}, we reproduce the positive outcomes of eigenvalue mirroring for such scenarios while clamping produces artifacts.
As suggested by the previous experiment, PDN and PPN face challenges in this inertia-free setting, indicating that unconditional projection might be preferred in this setting.

% Figure: Quasistatics

\begin{figure}
    \centering
    % ── big runtime bar plot ───────────────────────────────────────
    \includegraphics[width=\columnwidth]{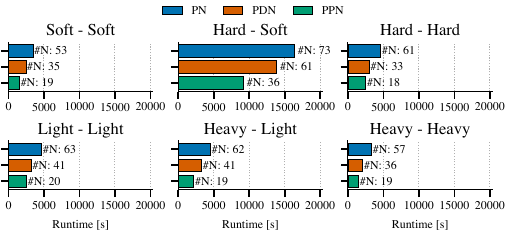}

    \medskip   % vertical gap between the plot and the snapshots
  
    % ── 2×2 snapshot grid with arrows + global labels (TikZ) ──────
    \begin{tikzpicture}[inner sep=0]
      \sffamily
      % 1. matrix = two columns × two rows of images
      \matrix (m) [matrix of nodes,
                   column sep=17pt,   % horizontal gap between columns
                   row sep=3pt]      % vertical gap between rows
      {
        |[name=SS0]|
          \includegraphics[width=0.375\columnwidth,
                           trim={0 350 0 350},clip]{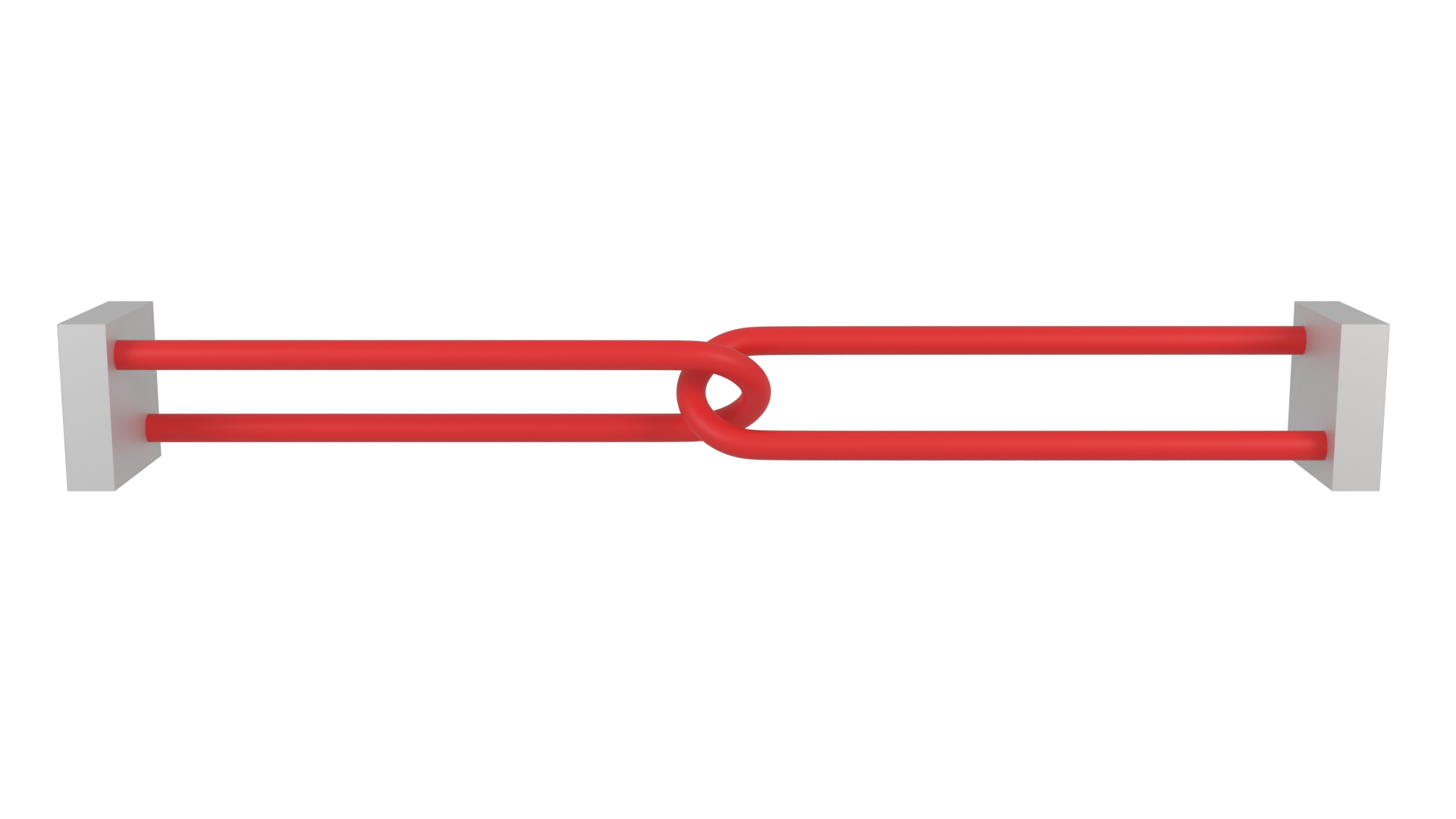} &
        |[name=SS1]|
          \includegraphics[width=0.375\columnwidth,
                           trim={0 350 0 350},clip]{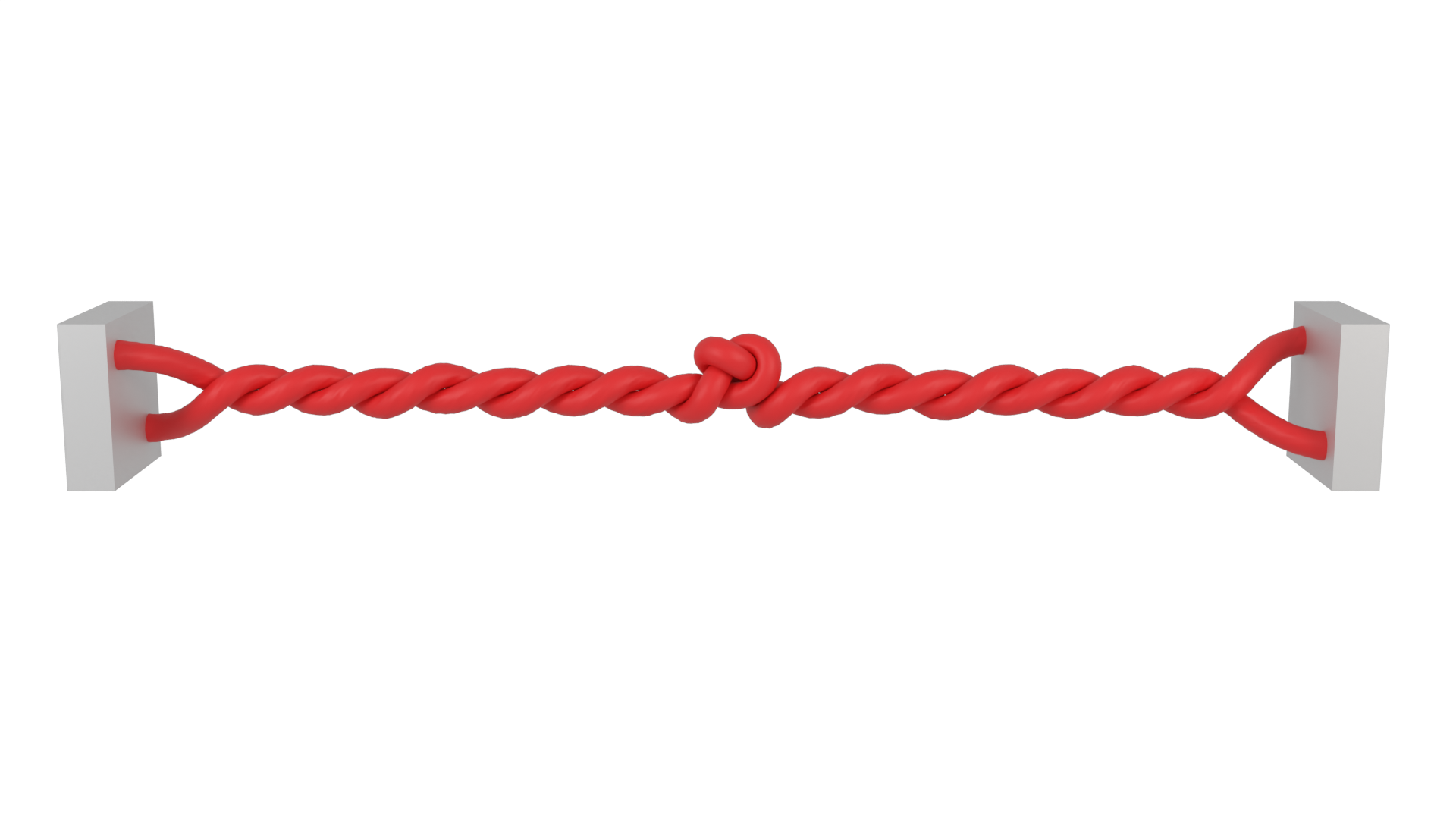} \\
        |[name=HS0]|
          \includegraphics[width=0.375\columnwidth,
                           trim={0 350 0 350},clip]{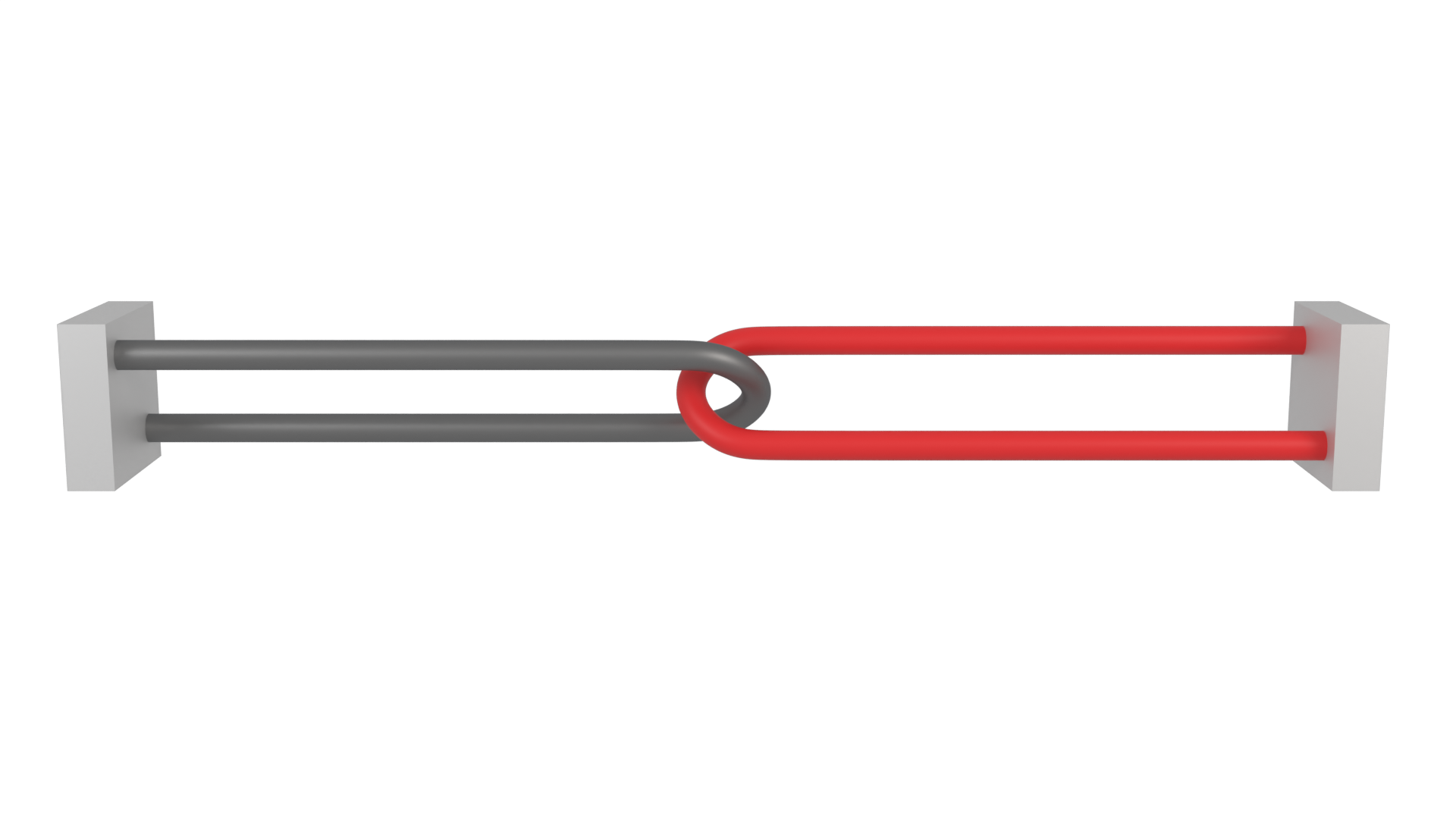} &
        |[name=HS1]|
          \includegraphics[width=0.375\columnwidth,
                           trim={0 350 0 350},clip]{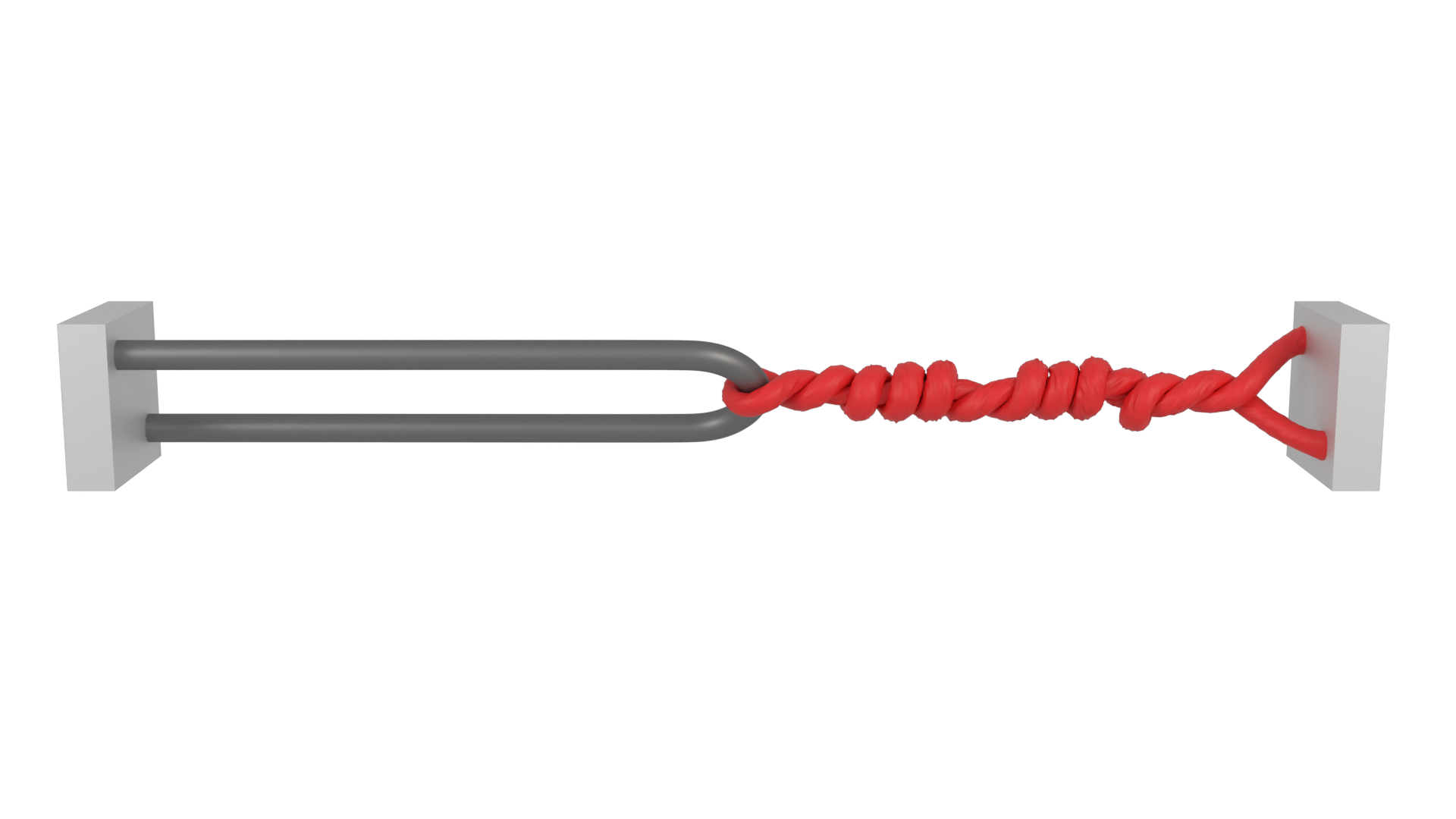} \\
      };
  
      % 2. arrows centred in each row
      \draw[imgarrow] (SS0.east) -- (SS1.west);
      \draw[imgarrow] (HS0.east) -- (HS1.west);
  
      % 3. global column labels
      \node[below=2pt of HS0.south, anchor=north] {\scriptsize Initial State};
      \node[below=2pt of HS1.south, anchor=north] {\scriptsize Final State};
  
      % 4. global row labels
      \node[left=6pt of SS0.west, anchor=east] {\scriptsize Soft -- Soft};
      \node[left=6pt of HS0.west, anchor=east] {\scriptsize Hard -- Soft};
    \end{tikzpicture}
  
    \caption{
        \textbf{U-Turn}.
        Two elastic cylinders are interlaced and twisted.
        Above, we compare solvers under varying Young's moduli (top) and densities (bottom).
        Below, the initial and final states are shown.
        All simulations except the ``Hard-Soft'' produce the same deformation.
        }
    \label{fig:rods}
  \end{figure}

\paragraph*{Large Ratios}
We compare all three solvers in simulations featuring large stiffness ($E = 10^{6}$--$10^{10}\,\mathrm{Pa}$) and density ($\rho = 10^{1}$--$10^{4}\,\mathrm{kg}\,\mathrm{m}^{-3}$) ratios in Fig.~\ref{fig:rods}.
Here, PPN consistently yields both the fewest Newton iterations and the fastest runtimes.
While on average PPN projects 6.2\% of the element Hessians, PDN projects 88.4\%.
On average, PPN requires only 35.6\% and 54.5\% of the Newton iterations of PN and PDN, respectively, corresponding to speedups of $\times$2.02 and $\times$1.42.
These findings suggest that the residual-based heuristic in PPN remains robust even when adjacent elements exhibit curvature variations spanning several orders of magnitude.

% Figure: Rods

\begin{figure}
    \centering
    \includegraphics[width=\columnwidth,trim={0 0 0 0},clip]{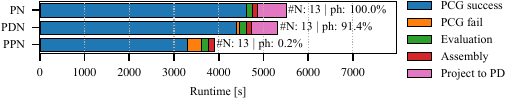}\\
    \vspace{0.5em}
    \includegraphics[width=\columnwidth,trim={0 80 0 300},clip]{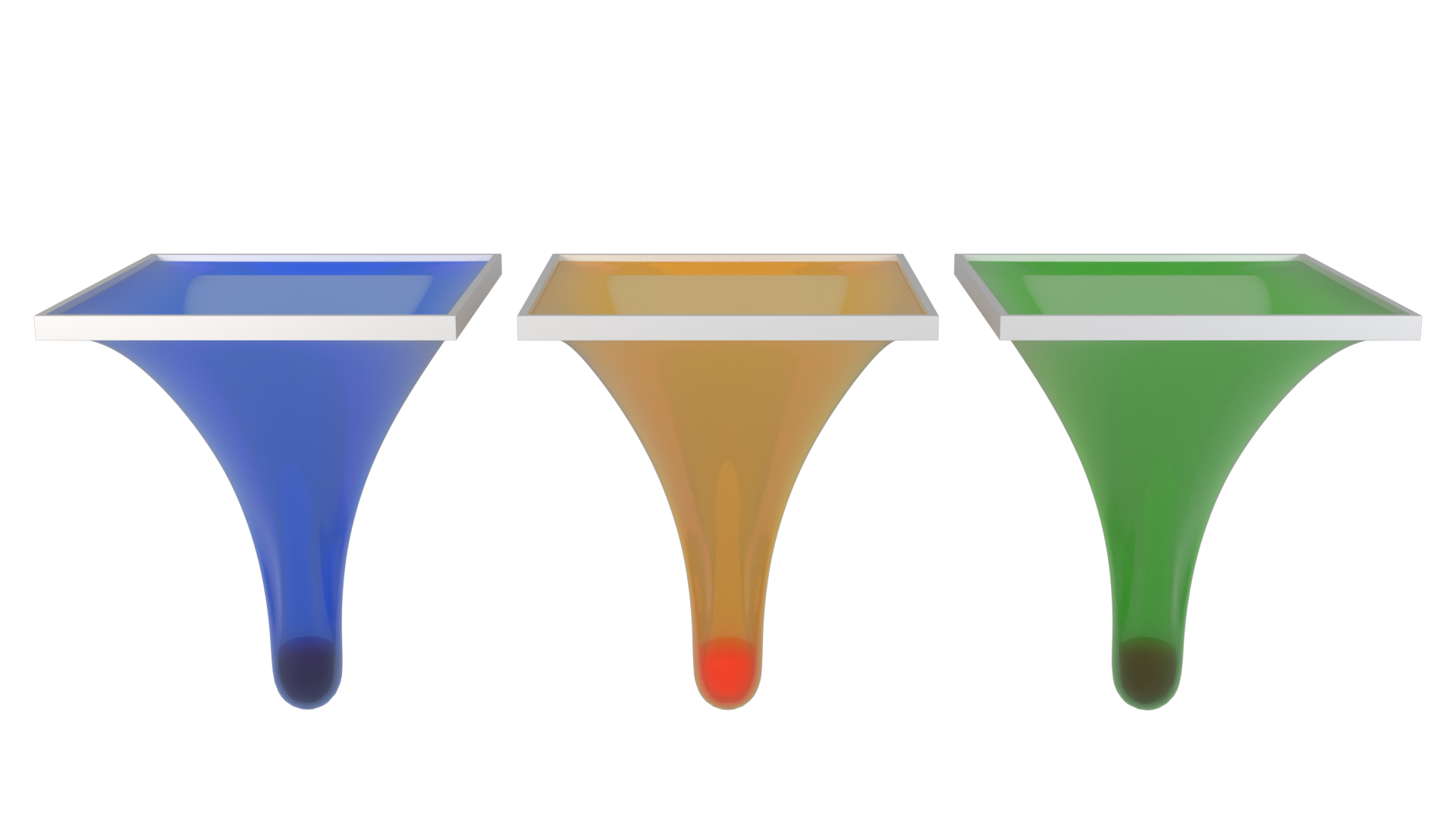}
    \caption{
        \ch{
            \textbf{Jelly Catch}.
            A heavy elastic ball falls on a thin sheet of very light elastic material clamped on its perimeter, resulting in very large strain in the material.
            Top: Comparison between the solvers.
            Bottom: Time step of maximum strain for PN, PDN and PPN, respectively from left to right.
        }
    }
    \label{fig:density_ratio}
\end{figure}

\ch{
    Finally, we quantify geometrical reproducibility across the three methods using a simulation characterized by large strains and extreme density ratios ($\rho = 10^{1}$--$10^{4}\,\mathrm{kg}\,\mathrm{m}^{-3}$), as shown in Fig.~\ref{fig:density_ratio}.
    The mean area-weighted surface distance over the full time sequence is just 0.34~$\mathrm{mm}$ (0.057\% of the domain size), with a maximum deviation of 1.04\%.
    This confirms that the choice of projection strategy does not compromise the fidelity of the outcome, yet PPN delivers speedups of $\times$1.39 and $\times$1.35 over PN and PDN, respectively.
}

% Figure: Density Ratio

% |     |   simulation_time |   ndofs |   runtime_actual_total |   n_total_newton |   total_ratio_projected_hessians |   n_total_line_search_iterations |   n_total_cg_iterations |
% |-----|-------------------|---------|------------------------|------------------|----------------------------------|----------------------------------|-------------------------|
% | PN  |             8.033 |  569745 |               5593.355 |         3272.000 |                            1.000 |                          885.000 |             3095416.000 |
% | PDN |             8.033 |  569745 |               5415.229 |         3317.000 |                            0.914 |                         1140.000 |             2978054.000 |
% | PPN |             8.033 |  569745 |               4015.687 |         3241.000 |                            0.002 |                         2545.000 |             2369078.000 |
% PDN vs PN -> Avg: 0.000135237 m | Weighted Avg: 0.000191876 m | Max: 0.00378716 m
% PPN vs PN -> Avg: 0.00036201 m | Weighted Avg: 0.000488863 m | Max: 0.00636461 m (1.04 %)
% Global Geometric Variation (Weighted Avg): 0.000340365 m (5.5693e-4 %, 0.057 %)
% Domain length 0.611137

\paragraph*{Contact-free}
The three solvers are compared in a contact-free simulation of an elastic armadillo in Fig.~\ref{fig:slingshot}.
Even in this simpler setting, PDN ends up projecting more than 50\% of all the element Hessians, while PPN only needs to project less than 3\%.
PPN reduces the Newton iterations by 53\% and 20\% with respect to PN and PDN, demonstrating that PPN's effectiveness is not exclusive to scenarios with complex frictional contact.
Corresponding speedups are $\times2.5$ and $\times1.34$.

% Figure: Armadillo Slingshot

%|     |   runtime_actual_total |   n_total_newton |   total_ratio_projected_hessians |
%|-----|------------------------|------------------|----------------------------------|
%| PN  |                420.041 |         7958.000 |                            1.000 |
%| PDN |                225.248 |         4515.000 |                            0.573 |
%| PPN |                168.726 |         3911.000 |                            0.026 |

\begin{figure}
    \centering
    \begin{tikzpicture}[inner sep=0] % no extra padding
        % top / overlay image
        \node[anchor=south west] (bg) at (0.39\columnwidth,0.32\columnwidth)
        {\includegraphics[width=0.370\columnwidth,trim={60 280 60 300},clip]{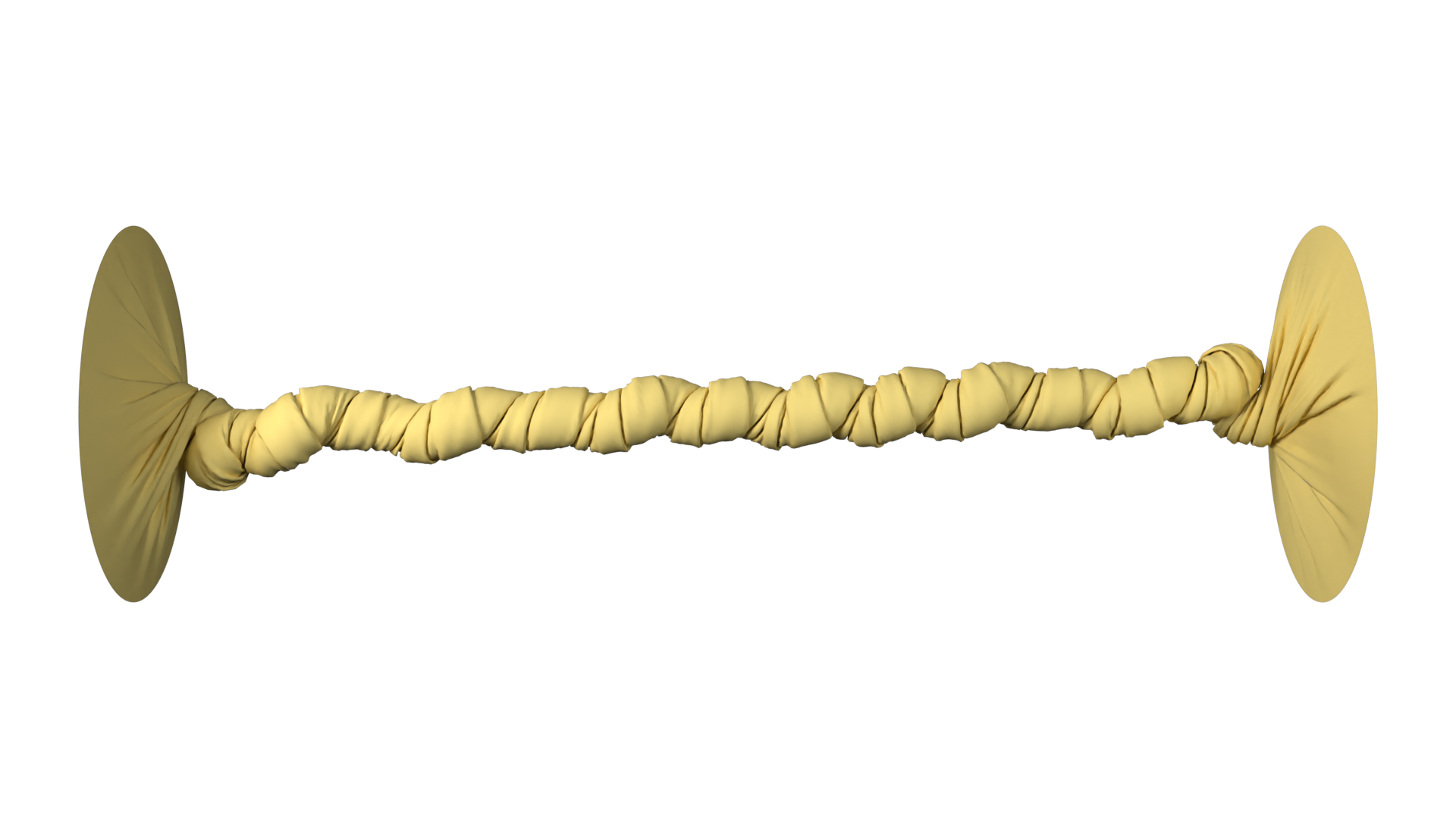}};

        % bottom / background image
        \node[anchor=south west] at (0,0)
        {\includegraphics[width=\columnwidth]{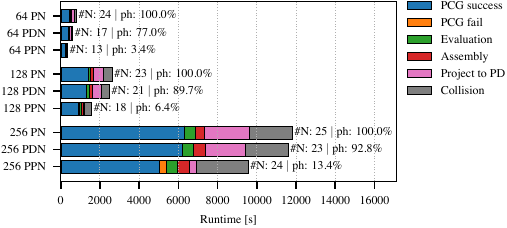}};
    \end{tikzpicture}
    \caption{
        \textbf{Twisting cloth}.
        A cloth cylinder is twisted by rotating its ends in opposite directions using three mesh resolutions.
        Final configuration of the finest resolution shown inside the plot.
        }
    \label{fig:twisting_cloth}
\end{figure}

\paragraph*{Codimensional}
%A cloth cylinder with different resolutions is twisted by rotating its ends in opposite directions.
We compare the three solvers in a contact-rich cloth simulation using three resolutions with vertex counts of $64^2$, $128^2$ and $256^2$ in Fig.~\ref{fig:twisting_cloth}.
In this challenging scenario, PPN provides significant improvements for the coarsest mesh: 
a reduction of 45.8\% and 30.7\% Newton iterations, and 55.5\% and 41.7\% of runtime in relation to PN and PDN, respectively.
For the finest discretization, PPN does not reduce iterations but still achieves more than 17\% performance improvement over the alternatives.

% Figure: Twisting cloth

% |         |   runtime_actual_total |   n_total_newton |   total_ratio_projected_hessians |
% |---------|------------------------|------------------|----------------------------------|
% | 64 PN   |                795.056 |        14571.000 |                            1.000 |
% | 64 PDN  |                602.563 |        10542.000 |                            0.770 |
% | 64 PPN  |                351.560 |         8036.000 |                            0.034 |
% | 128 PN  |               2617.533 |        13800.000 |                            1.000 |
% | 128 PDN |               2483.500 |        12977.000 |                            0.897 |
% | 128 PPN |               1569.462 |        11067.000 |                            0.064 |
% | 256 PN  |              11806.825 |        15198.000 |                            1.000 |
% | 256 PDN |              11592.695 |        14111.000 |                            0.928 |
% | 256 PPN |               9548.726 |        14939.000 |                            0.134 |

% avg speedup to PN 1.7219261148022593
% avg speedup to PDN 1.5034717171541827
% avg reduction n_newton PN 0.22119292580691233
% avg reduction n_newton PDN 0.10874055151061886

\begin{figure}
    \centering
    \includegraphics[width=\columnwidth,trim={0 0 0 0},clip]{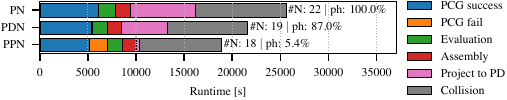}\\
    \includegraphics[height=0.4\columnwidth,trim={800 30 800 50},clip]{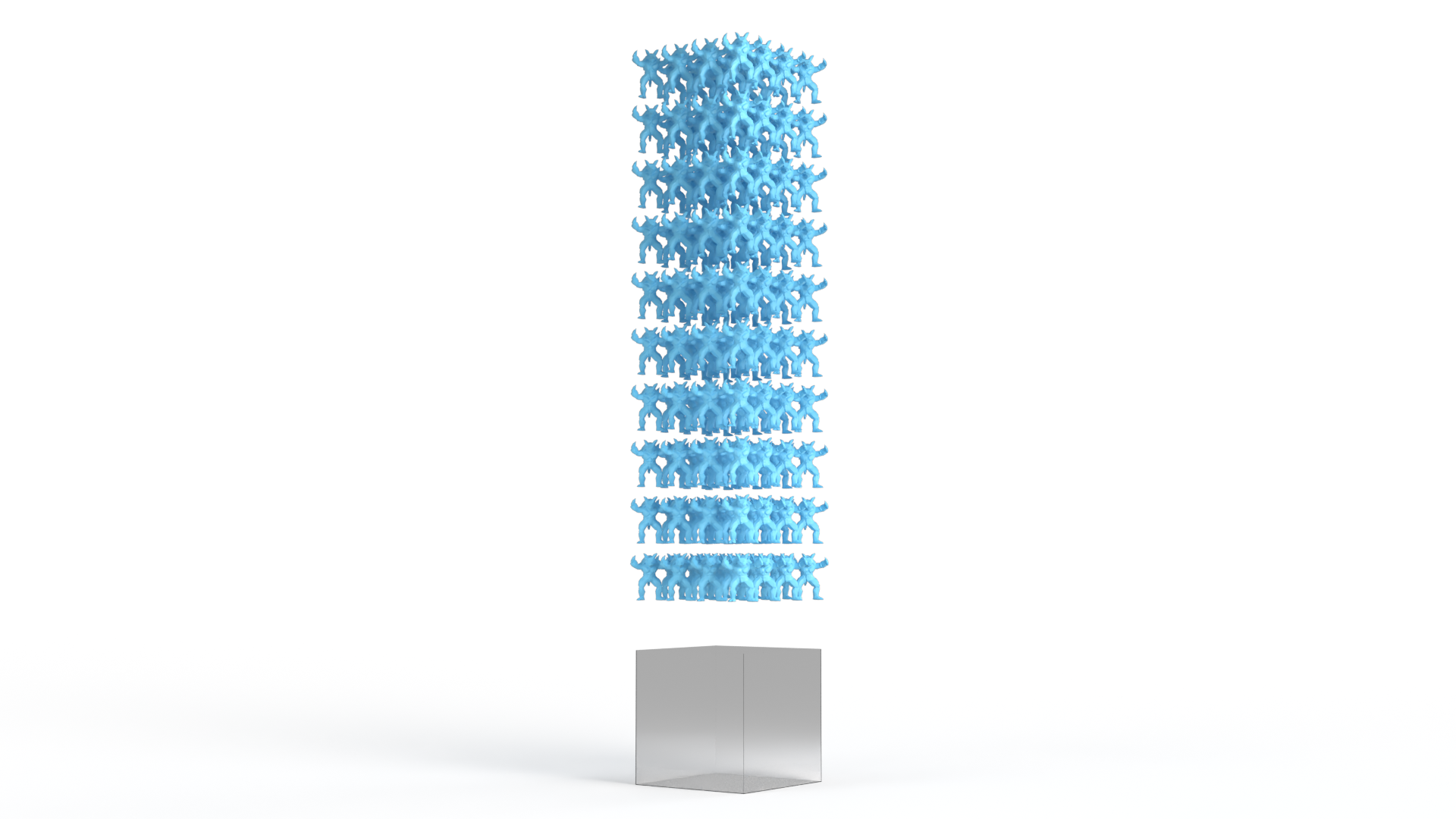}
    \includegraphics[height=0.4\columnwidth,trim={450 0 450 200},clip]{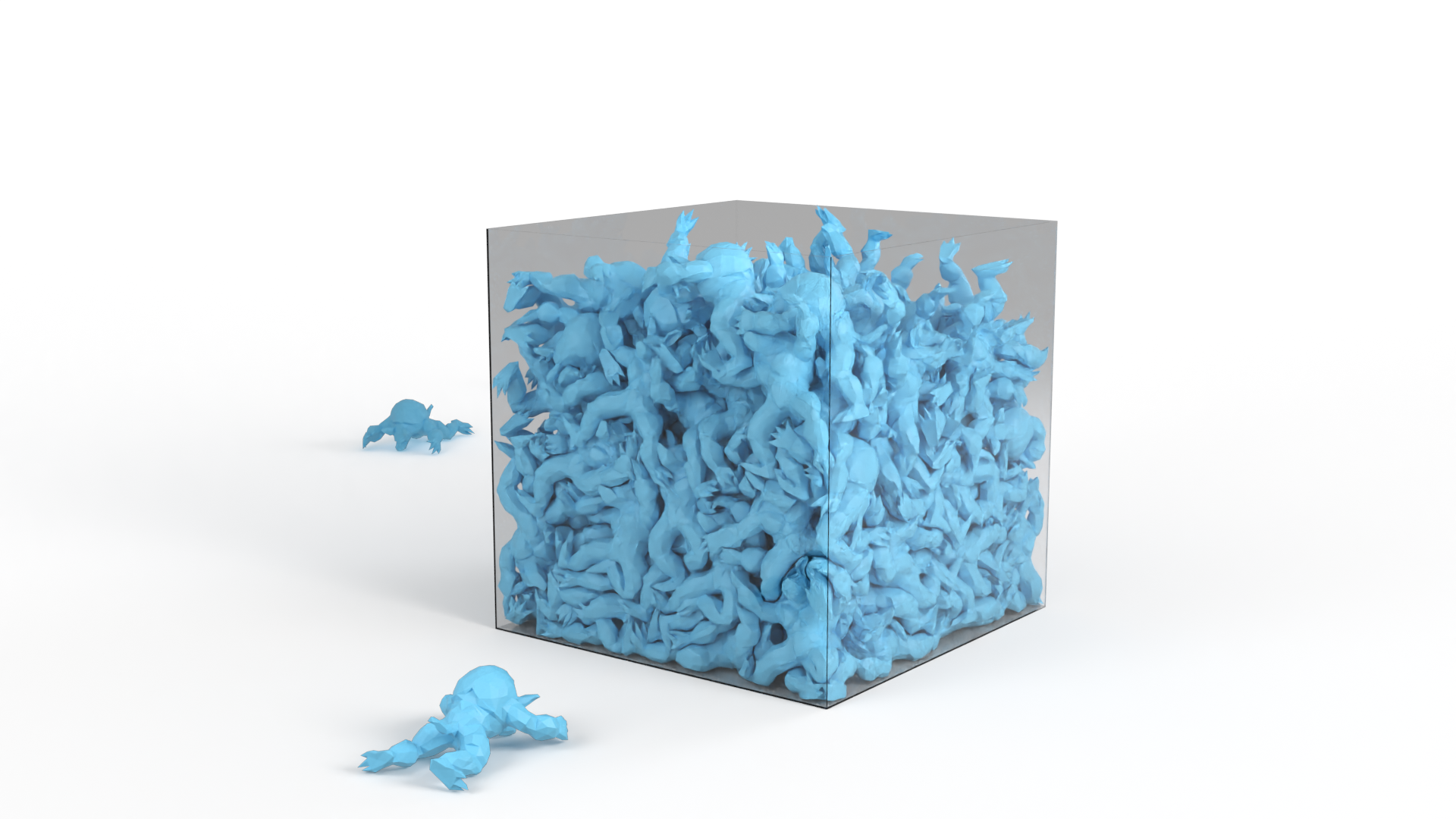}
    \caption{
        \textbf{Armadillo drop}.
        160 elastic armadillos are dropped into a rigid box.
        Comparison between solvers on top.
        Initial and final state below.
        }
    \label{fig:drop}
\end{figure}

\begin{figure}
    \centering
    \includegraphics[width=\columnwidth,trim={0 0 0 0},clip]{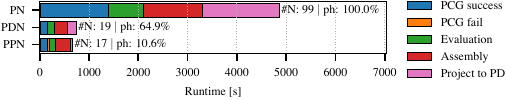}\\
    \vspace{0.5em}
    \includegraphics[width=0.75\columnwidth,trim={0 40 0 0},clip]{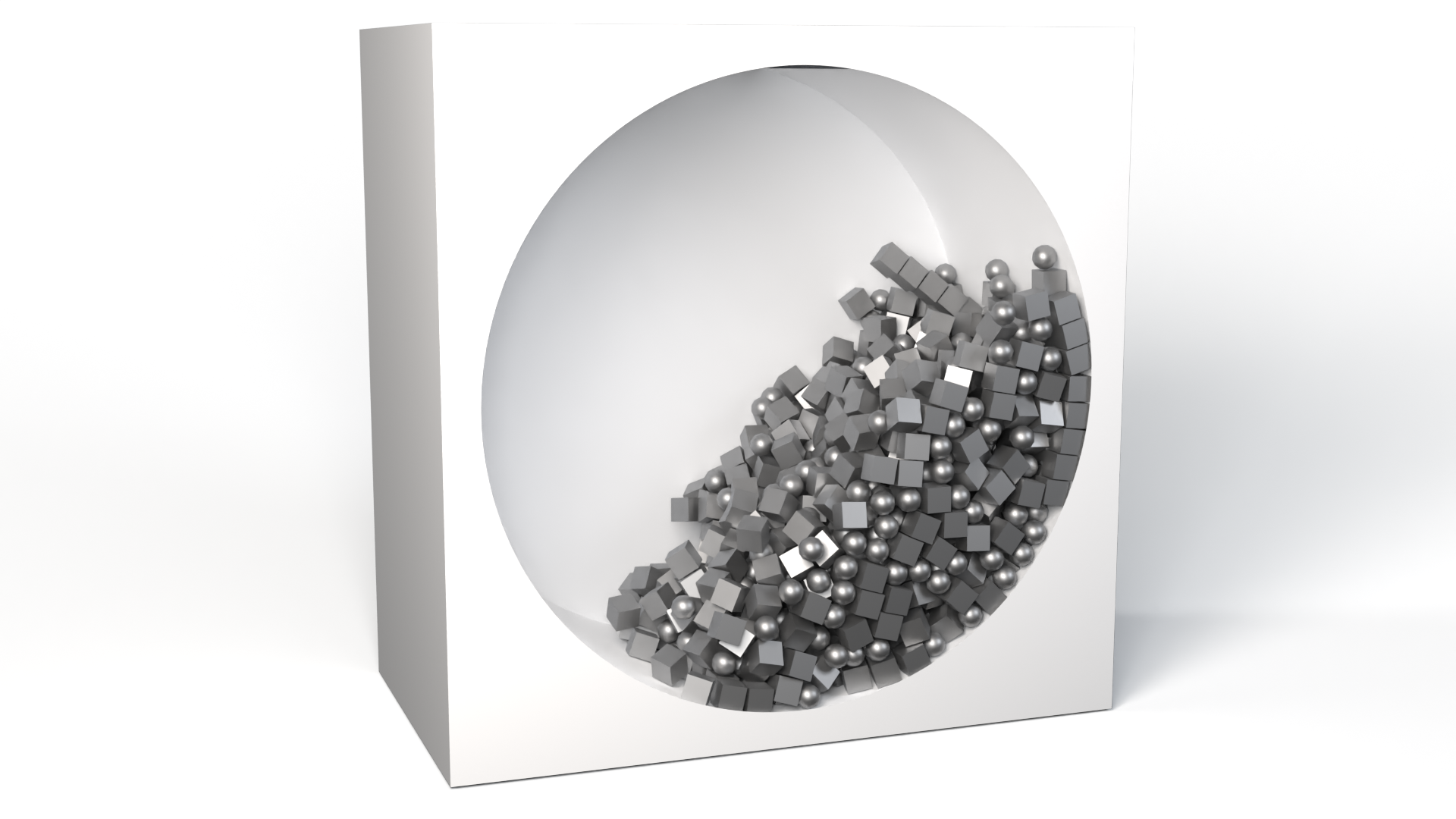}
    \caption{
        \textbf{Tumbler}.
        More than 1000 rigid bodies collide inside a spinning tumbler.
        Comparison between solver on top.
        Collision runtime is omitted for clarity, as it dominates total cost.
        An intermediate state is shown below.
        }
    \label{fig:tumbler}
\end{figure}

\paragraph*{Impact-rich}
We test two scenes featuring high-energy impacts in Fig.~\ref{fig:drop} and~\ref{fig:tumbler}.
The former simulates elastic objects and the latter rigid bodies.
To preserve vividness, these simulations use a time step of $1/300\,$\SI{}{\second}, as larger time steps resulted in visibly damped dynamics.
%160 elastic armadillos are dropped into a rigid box in Fig.~\ref{fig:drop}, and more than 1000 rigid bodies collide inside a tumbler in Fig.~\ref{fig:tumbler}.
Although PPN still greatly reduces the number of projections, its advantage over PDN in terms of Newton iterations is more modest.
Nevertheless, runtime was reduced by 26.2\% and 12.6\% in relation to PN and PDN for elastic scene, and by 86.5\% and 10.4\% for the rigid body one.
Notably, PN struggles significantly in the rigid body scene, requiring more than five times as many iterations and representing an outlier in our tests.

% Figure: Drop

% |     |   runtime_actual_total |   n_total_newton |   total_ratio_projected_hessians |
% |-----|------------------------|------------------|----------------------------------|
% | PN  |              25561.949 |        50624.000 |                            1.000 |
% | PDN |              21598.504 |        44740.000 |                            0.870 |
% | PPN |              18872.698 |        42755.000 |                            0.054 |

% Figure: Tumbler

% |     |   runtime_actual_total |   n_total_newton |   total_ratio_projected_hessians |
% |-----|------------------------|------------------|----------------------------------|
% | PN  |                 4854.0 |       446771.000 |                            1.000 |
% | PDN |                  733.4 |        88247.000 |                            0.649 |
% | PPN |                  657.4 |        77462.000 |                            0.106 |

% Overall results
% | Scene               |   % proj avoided |   % Newton iterations reduction |   speedup |
% |---------------------|------------------|---------------------------------|-----------|
% | Press               |               92 |                              14 |     1.540 |
% | U-Turn              |               94 |                              41 |     1.420 |
% | Armadillo slingshot |               97 |                              20 |     1.340 |
% | Twisting cloth      |               90 |                              11 |     1.500 |
% | Armadillo drop      |               94 |                               4 |     1.140 |
% | Tumbler             |               90 |                              12 |     1.120 |

\begin{table*}[t]
  \centering
  \caption{
    Scene parameters.
    All simulation use \SI{0.5}{\milli\meter} IPC contact distance.
  }
  \label{tab:sim_params}
  \begin{tabular}{lcccccc}
    \toprule
    \textbf{Scene} &
    $\boldsymbol{n_{\mathrm{dof}}}$\! &
    $\boldsymbol{\Delta t}\;[\mathrm{s}]$ &
    \textbf{Dimension}\,[m] &
    \textbf{Duration}\,[s] &
    \textbf{Material $(E,\nu)$} &
    \textbf{Density} \\
    \midrule
    Rolling sphere         & $3\,$K   & $1/30$ & $0.15$                         & $12$ & $1\times10^{3}\,\mathrm{Pa},\;0.49$ & $1000\;\mathrm{kg\,m^{-3}}$ \\
    Press                  & $47\,$K  & $1/30$ & $0.30$                         & $5$ & $1\times10^{5}\,\mathrm{Pa},\;0.40$ & $1000\;\mathrm{kg\,m^{-3}}$ \\
    Quasistatic extrusion & variable & $\infty$ & $0.50$                       & -- & $1\times10^{8}\,\mathrm{Pa},\;0.49$ & -- \\
    U-Turn                 & $87\,$K  & $1/30$ & $1.85$                         & $12$ & variable,\;$0.49$                    & variable \\
    Jelly Catch            & $570\,$K  & $1/30$ & $0.5$                         & $10$ & variable,\;$0.49$                    & variable \\
    Armadillo slingshot    & $71\,$K  & $1/30$ & $1.00$                         & $15$ & $1\times10^{5}\,\mathrm{Pa},\;0.40$ & $1000\;\mathrm{kg\,m^{-3}}$ \\
    Twisting cloth         & variable & $1/30$ & $0.50,$ $0.001$ thick          & $20$ & $1\times10^{5}\,\mathrm{Pa},\;0.30$ & $0.20\;\mathrm{kg\,m^{-2}}$ \\
    Armadillo drop         & $566\,$K & $1/300$& $0.35$                         & $7.5$ & $1\times10^{4}\,\mathrm{Pa},\;0.45$ & $1000\;\mathrm{kg\,m^{-3}}$ \\
    Tumbler                & $7\,$K   & $1/300$& $0.50$                         & $15$ & rigid                                & $1000\;\mathrm{kg\,m^{-3}}$ \\
    \bottomrule
  \end{tabular}
\end{table*}

\section{Limitations And Future Work}
\label{sec:limitations}

While PPN reliably reduces element projections and Newton iterations in any simulator, the net speedup can vary depending on solver choices (e.g., direct vs.\ iterative) and execution platforms (e.g., CPU vs.\ GPU).
Codebases that incorporate analytic projections may see lower gains than those relying on numerical eigendecompositions, and applications using a fixed number of Newton iterations may only benefit from the reduced projections.

Our results also indicate that current on-demand projection strategies, including PDN and PPN, perform poorly in scenarios lacking strong mass matrix contributions (or alternative forms of regularization), as in very large time steps or in the quasistatic limit.
However, the time steps where this becomes a problem are rare in practice due to the associated numerical damping and loss of detail.
In any case, for realistic choices of time step sizes, including relatively large ones such as $1/30\,$\si{\second} PPN consistently excels.

% Finally, our residual-based heuristic does not require additional calculations, so there is no overhead over Newton's Method if no projections are needed.
% While this heuristic is shown to be highly effective, other criteria more directly tied to local assembled indefiniteness may further improve convergence.  
% We will explore these possibilities in future research.  

\section{Conclusion}
\label{sec:conclusions}

We introduced Progressively Projected Newton, a direct replacement for Projected Newton \ch{in dynamic simulations} that guarantees descent directions while reducing element projections by an order of magnitude and improving convergence.
% Our method can be easily integrated into existing PN-based simulators.
PPN begins each Newton iteration with the unmodified Hessian and only projects elements incrementally when the linear solver detects indefiniteness, guided by a residual-driven tolerance that adapts across iterations.

Extensive experiments on dynamic simulations of deformable solids, shells, frictional contact, and rigid bodies demonstrate that PPN consistently performs 90\% fewer projections, decreases the number of Newton iterations by up to 50\% compared to PN, and achieves total simulation speedups of up to $\times 2.5$ relative to PN and up to $\times 1.5$ over PDN.
Hence, PPN stands out as the most efficient method among the three for robustly handling indefiniteness in second-order optimization time integration.

%-------------------------------------------------------------------------
%% Acknowledgements
\section*{Acknowledgements}
  This work is funded by the Deutsche Forschungsgemeinschaft (DFG, German Research Foundation) --- project number 281466253. Open Access funding enabled and organized by Projekt DEAL.

%-------------------------------------------------------------------------
% bibtex
\bibliographystyle{eg-alpha-doi} 
\bibliography{bibliography}

% biblatex with biber
% \printbibliography                

\end{document}